\documentclass[twocolumn,english,prl,notitlepage,superscriptaddress,nobibnotes,nolongbibliography]{revtex4-2}
\usepackage{graphicx}
\usepackage{amsmath}
\usepackage{amssymb}
\usepackage{bm}
\usepackage{enumerate}
\usepackage{xfrac}
\usepackage[caption=false]{subfig}
\usepackage[normalem]{ulem}
\usepackage[usenames,dvipsnames]{xcolor}
\usepackage{txfonts}
\usepackage[mathscr]{euscript}
\usepackage[pdftex,
		  pdflang={en-US},
            colorlinks=true,
            citecolor=blue,
            linkcolor=blue,
            urlcolor=blue,
            pdfauthor={},
	      pdftitle={Pseudogap_HeavyFermions}]{hyperref}
\usepackage{verbatim}
\usepackage{ulem}
\begin{document}

\title{Excess heat capacity in magnetically ordered Ce heavy fermion metals}
\author{A. Scheie}
\email{scheie@lanl.gov} 
\affiliation{MPA-Q, Los Alamos National Laboratory, Los Alamos, NM 87545, USA}

\author{Yu Liu}  
\affiliation{MPA-Q, Los Alamos National Laboratory, Los Alamos, NM 87545, USA}

\author{E. A. Ghioldi}  
\affiliation{Department of Physics and Astronomy, University of Tennessee, Knoxville, TN, USA}

\author{S. Fender}  
\affiliation{MPA-Q, Los Alamos National Laboratory, Los Alamos, NM 87545, USA}

\author{P. F. S. Rosa}  
\affiliation{MPA-Q, Los Alamos National Laboratory, Los Alamos, NM 87545, USA}

\author{E. D. Bauer}  
\affiliation{MPA-Q, Los Alamos National Laboratory, Los Alamos, NM 87545, USA}


\author{Jian-Xin Zhu} 
\affiliation{Theoretical Division and Center for Integrated Nanotechnologies, Los Alamos National Laboratory, Los Alamos, NM 87545, USA}

\author{F. Ronning} 
\email{fronning@lanl.gov} 
\affiliation{MPA-Q, Los Alamos National Laboratory, Los Alamos, NM 87545, USA}

\date{\today}


\begin{abstract}
We study the magnetic heat capacity of a series of magnetically ordered Ce-based heavy fermion materials, which show an anomalous $T^3$ heat capacity in excess of the phonon contribution in many materials. For compounds for which magnon models have been worked out, we show that the local-moment magnon heat capacity derived from the measured magnon spectra underestimates the experimental specific heat. The excess heat capacity reveals increasing density of states with increasing energy, akin to a pseudogap. We show that this anomalous temperature-dependent term is not associated with proximity to a quantum critical point (QCP), but is strongly correlated with $T_N$, indicating the anomalous excitations are governed by the magnetic exchange interaction. This insight may hold key information for understanding magnetically ordered heavy fermions. 
\end{abstract}
\maketitle


First discovered in the 1970's \cite{Andres_1975,Steglich_1979}, heavy fermions are a prototypical problem of strongly correlated electron systems \cite{Steward_1984_RMP,fisk1995physics,pavarini2015many,Wirth2016,Steglich_2016,shaginyan2022peculiar}.
Deriving their name from an anomalously large effective electron mass at low temperatures, these materials display a variety of strongly correlated quantum phases, including 
non-Fermi liquids \cite{si2010heavy,shaginyan2022peculiar},
unconventional superconductivity \cite{Steglich_2016}, volume collapse \cite{LAVAGNA1982210}, topological Kondo insulators \cite{Dzero_2010}, and hidden order \cite{Mydosh_2014}. Many share similar phenomenology of quantum criticality, summarized by the famous Doniach phase diagram \cite{Doniach1977}. 
It is known that heavy fermion behavior arises from the interactions between local and itinerant electrons. Yet, despite decades of work, there is no microscopic model able to account for their behavior. 
This signifies key gaps in our understanding of superconductivity, non-quasiparticle transport, and fundamental many-body quantum physics. 

As the list of heavy fermion materials continues to grow but theory is still lacking, one route to explaining heavy fermions is looking for trends across materials families \cite{KADOWAKI1986507,Wilson_1975}. In this paper, we focus on heat capacity of magnetically ordered Ce heavy fermion materials \cite{Weng_2016}. 
Beginning with CeIn$_3$, we show that a common feature of these compounds is an anomalous density of states at low energies (in addition to $T$-linear Sommerfeld coefficient), often taking the form of an approximate $T^3$ term in heat capacity.  In certain cases, where a rigorous magnon model has been worked out, we show that the experimental heat capacity far exceeds the bosonic magnon heat capacity at low temperatures. Correlation analysis shows this density of states to be uncorrelated with proximity to quantum criticality but strongly correlated with the ordering temperature. Thus, these anomalous excitations are related to the magnetic exchange interaction. 

\begin{figure}
	\centering
	\includegraphics[width=0.47\textwidth]{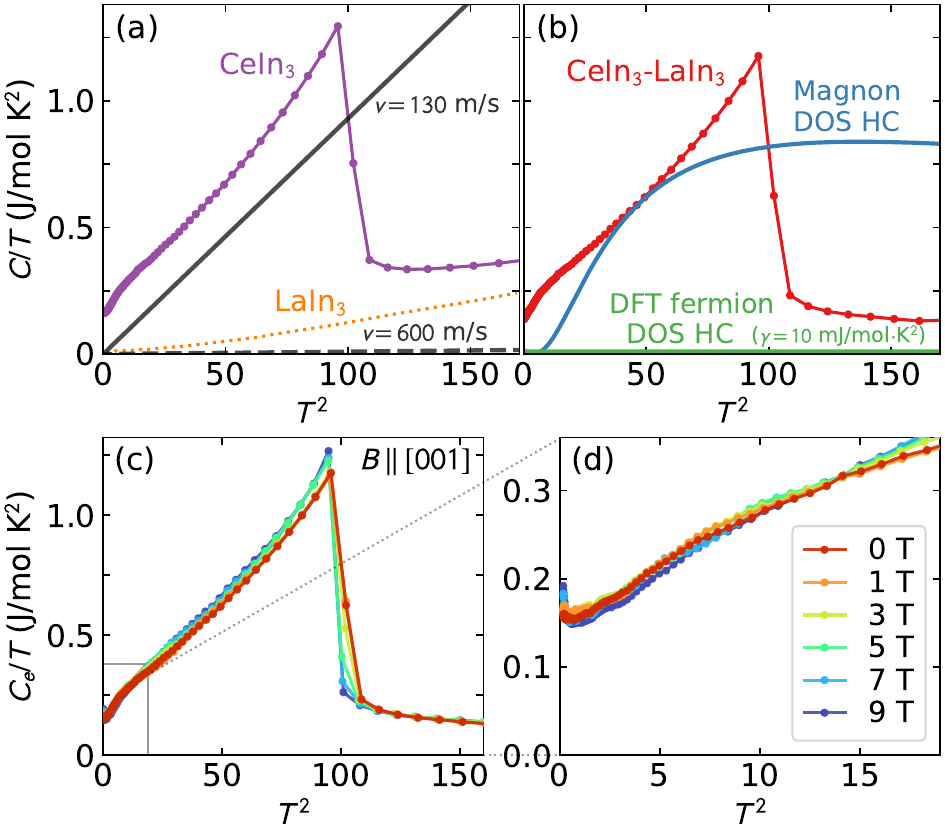}
	\caption{CeIn$_3$ heat capacity compared to magnon models. Panel (a) shows the simplistic heat capacity from Eq. \ref{eq:LinearMagnonHC}, with the low-energy velocity from the neutron-derived magnon model $v\approx 600$~m/s \cite{simeth2022microscopic}. This is two orders of magnitude smaller than experiment. Also shown in heat capacity of nonmagnetic analogue LaIn$_3$ from Ref. \cite{Berry_2010}. Panel (b) shows the calculated heat capacity from a more sophisticated model, integrating over the whole Brillouin zone for the magnon band structure (blue) and the electron band structure (green). Both significantly underestimate the low-energy density of states compared to the electronic specific heat $C_e$. Panel (c) shows the $[001]$ field dependent heat capacity, with (d) as a closer view of the low-$T$ behavior. Application of a 9~T field makes almost no difference to the $T^3$ heat capacity, contrary to expected magnon behavior.}
	\label{fig:CeIn3}
\end{figure}

By way of introduction, let us begin by 
examining the heat capacity of the magnetically ordered heavy fermion system CeIn$_3$. This compound magnetically orders at $T_N = 10.23(1)$~K \cite{Lawrence_1980}, and has superconducting and non-Fermi liquid properties under pressure \cite{Mathur1998Nature,Knebel_2001}. 
We measured its heat capacity using a Quantum Design Physical Property Measurement System (PPMS) from 0.4~K to  20~K that utilizes a quasi-adiabatic thermal relaxation method, and the data is shown in Fig. \ref{fig:CeIn3}. 
Plotting the data on a $C/T$ vs $T^2$ graph shows a nearly straight line below $T_N$, indicating $T^3$ heat capacity (slope) with a $T$-linear term ($y$-axis offset). 
The $T$-linear term is explicable (at least phenomenologically) as a Sommerfeld term from enhanced fermion mass \cite{AshcroftMermin}. The $T^3$ term, however, is much larger than the phonon heat capacity (approximated by the nonmagnetic LaIn$_3$ \cite{Berry_2010}) and is more of a challenge. 


In theory, gapless linear dispersive magnons in three dimensions in the low-temperature limit  give $T^3$ heat capacity 
\begin{equation}
    c_{mol} =  N_A v_0 \frac{4\pi^2}{15} k_B \bigg( \frac{k_B T}{h v} \bigg)^3
    \label{eq:LinearMagnonHC}
\end{equation}
where $v_0$ is the volume of the unit cell, $h$ is Planck's constant, and $v$ is the velocity of the modes \cite{AshcroftMermin}. Recent CeIn$_3$ neutron scattering studies have shown gapless linear dispersive magnons with a velocity of $v \approx 600$~m/s \cite{simeth2022microscopic}. However, the calculated heat capacity from such modes via Eq. \ref{eq:LinearMagnonHC}, shown in Fig. \ref{fig:CeIn3}(a), underestimates the specific heat by two orders of magnitude. (The slope of heat capacity suggests a magnon velocity $v \approx 130$~m/s, inconsistent with the neutron results.)  
We can improve this calculation by instead integrating over the full magnon band structure rather than just the bottom of the dispersion. Taking the CeIn$_3$ magnon dispersion from Ref.  \cite{simeth2022microscopic}, one can more rigorously calculate the heat capacity by numerically integrating over the entire Brillouin zone, 
\begin{equation}
    c_v = k_B \sum_s \int dk  \Big( \frac{\hbar \omega_s (k)}{k_B T} \Big)^2 \frac{e^{\hbar \omega_s (k) / k_B T}}{(e^{\hbar \omega_s (k) / k_B T} -1)^2}
    \label{eq:BosonicHC}
\end{equation}
summing over $s$ magnon modes where $\omega_s (k)$ are the mode dispersions \cite{AshcroftMermin}. The result of these calculations are shown in Fig. \ref{fig:CeIn3}(b), where $C_e$ is the electronic (phonon-subtracted) specific heat. The calculated heat capacity comes close to 
the $C_e(T)/T$ data 
near 5~K
(where the validity of theory is questionable: the expansion is only valid when the moment size is near saturation, likely $T \lesssim \frac{T_N}{2}$), but the  calculated heat capacity is far too small below $\sim 3$~K. Clearly, the anomalously large $T^3$ heat capacity cannot be explained by the derived local-moment magnon model.

Further evidence against the $T^3$ heat capacity being magnons is found in the field-dependent data, shown in Fig. \ref{fig:CeIn3}(c)-(d). Ordinarily, a magnetic field shifts magnon bands up in energy, decreasing the low energy density of states and suppressing the low-temperature Eq. \ref{eq:BosonicHC} heat capacity. However, the heat capacity below $T_N$ is barely affected by a magnetic field, indicating this density of states is not from local moment magnons. 

As a final attempt to explain the CeIn$_3$ $T^3$ heat capacity, we calculate the electron band structure with density functional theory (DFT). 
Using the CeIn$_3$ experimental crystal structure, we performed DFT calculations by using a full-potential linearized augmented plane wave (FP-LAPW) as implemented in the WIEN2k code~\cite{PBlaha:2001}. On top of the generalized gradient approximation (GGA)~\cite{Perdew_1996} for the exchange-correlation functional, we used a value of Hubbard $U_{eff}=6.0$ eV on Ce-4f electrons for a G-type antiferromagnetically ordered state with the magnetization imposed along (111) direction. The spin-orbit coupling was included in a second variational way. A plane wave cutoff $RK_{\text{max}}=8$ were taken with a $12\times 12 \times 12$ $\mathbf{k}$-points. 
The resulting heat capacity, calculated via Eq. \ref{eq:BosonicHC} but with fermionic statistics, are shown as the green line in Fig. \ref{fig:CeIn3}(b). Not only does it vastly underestimate the Sommerfeld $\gamma$ term ($\gamma = 9.88$~mJ/mol$\cdot$K$^2$), 
it has virtually no $T^3$ dependence with a $T^3$ prefactor $2.37(3) \times 10^{-5}$~J/mol$\cdot$K$^4$ for $T^2 < 50$~K$^2$, six orders of magnitude smaller than CeIn$_3$'s fitted $T^3$ prefactor $10.31(11)$~J/mol$\cdot$K$^4$. 
(If we renormalize the DFT band structure energy to yield larger DOS near the Fermi energy and match the empirical $\gamma = 130$~mJ/mol$\cdot$K$^2$, this still falls far short with $T^3$ prefactor $1.76(2) \times 10^{-2}$~J/mol$\cdot$K$^4$ for $T^2 < 10$~K$^2$, still three orders of magnitude too small.)
Thus, DFT electronic band structures are unable to explain the $T^3$ heat capacity. This is not so surprising, as DFT often struggles to capture strong correlations between electrons. 

Clearly, there is some significant density of states at low energy that pure magnon and pure electron band theory fails to capture. The strong correlations in CeIn$_3$ produce a substantial energy-dependent density of states (i.e., $T^3$ specific heat), not merely an enhanced electron mass (which would give $T$-linear specific heat). 


\begin{figure*}
	\centering
	\includegraphics[width=0.99\textwidth]{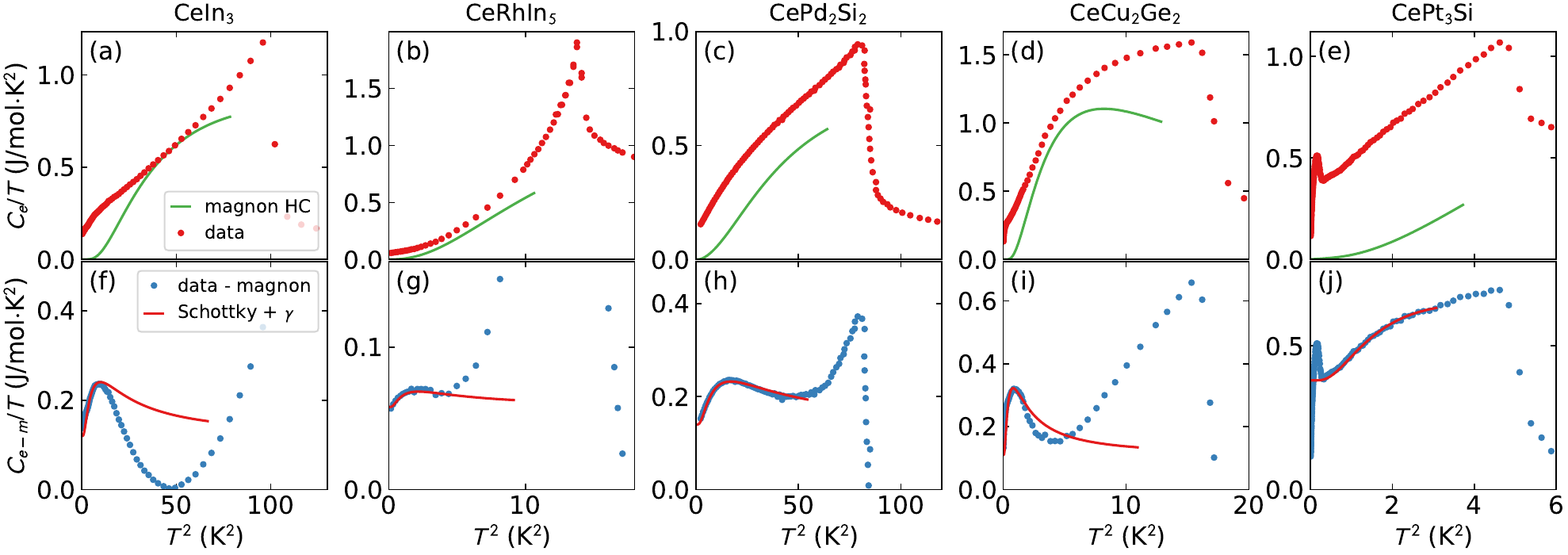}
	\caption{Electronic heat capacity of five different magnetically ordered heavy fermion materials for which a magnon model is available. (a)-(e) shows the raw data, which has had the lattice contribution subtracted (red) compared to the calculated magnon heat capacity (green). The bottom row (f)-(j) shows the data with the magnon model subtracted ($C_{e-m}$), compared to a Schottky anomaly offset by a $\gamma$ term. 
    In each case the extra density of states has the character of a (pseudo)gapped density of states. Contrast this with the Kondo effect heat capacity, where $C/T$ monotonically decreases with temperature \cite{DESGRANGES1982}.}
	\label{fig:HCmagnons}
\end{figure*}

Having observed such behavior in one magnetically ordered heavy fermion material, a natural question is how general is this behavior. 
In Fig.  \ref{fig:HCmagnons}, we compare experimental lattice-subtracted heat capacity to magnon heat capacity for five magnetically ordered heavy fermion materials for which a magnon model exists: CeRhIn$_5$ \cite{Das_2014_CeRhIn5}, CePd$_2$Si$_2$ \cite{Dijk_2000}, CeCu$_2$Ge$_2$ \cite{Knopp1989}, and CePt$_3$Si \cite{Faak_2008}. 
The data from these compounds, and the calculated magnon specific heat (Eq. \ref{eq:BosonicHC}), are shown in Fig. \ref{fig:HCmagnons}(a)-(e).  (For CeCu$_2$Ge$_2$ a proper magnon model does not exist, and the calculated magnon heat capacity is from an Einstein mode with the energy of the flat band measured in Ref. \cite{Knopp1989}.)

In every case, there is a large temperature-dependent specific heat term in the experimental data that can be not accounted for by the magnon model.  This is made more evident by the lower row (f)-(j) of Fig. \ref{fig:HCmagnons}, where the magnon calculated specific heat has been subtracted from the experimental heat capacity. In all compounds, the residual specific heat has a peak at low temperatures, which vaguely resembles a Schottky anomaly (indicated by the red lines). This is true even for CeRhIn$_5$, which has the smallest $\gamma$ value of the five compounds. This suggests some kind of (pseudo)gap in the density of states, wherein the density of states increases with increasing energy. 
In each compound, the excess heat capacity rises to 20-50\% of the  $\gamma$ value ($T \rightarrow 0$~K), by no means a small contribution. 
Furthermore, the pseudogap energy is consistently the same order as $T_N$ (see the Supplemental Materials \cite{SuppMat}), suggesting an energy scale governed by the magnetic order. 


The five compounds in Fig. \ref{fig:HCmagnons} had the luxury of a fitted magnon model, but we gain more insight by extending this analysis to a broader set of compounds. 
In Table \ref{tab:compounds} we examine 17 different magnetically ordered Ce-based heavy fermion materials. Taking their heat capacities below $T_N$ from the literature, we fit the lowest temperature data to 
\begin{equation}
    c = \gamma T + N_a k_B \bigg(\frac{T}{T_{\beta}} \bigg)^3
    \label{eq:HCfit}
\end{equation}
where $N_a$ is Avagadro's number and $T_{\beta}$ serves as a magnetic analogue of the Debye temperature. 
(This is not meant to imply that the true non-magnon specific heat is $T^3$ over many decades, but is meant to capture the lowest temperature behavior which, as Fig. \ref{fig:HCmagnons} shows, is mainly preserved when the local-moment magnon contribution is subtracted.)
The fits are shown in the Supplemental Materials \cite{SuppMat}, and the fitted values are listed in Table \ref{tab:compounds}.

\begin{table}
\caption{Experimental properties of various magnetically ordered Ce heavy fermion materials. $\gamma$ (Sommerfeld coefficient), $P_c$ (critical pressure), and $T_N$ (Neel temperature) are taken from the literature, but $T_{\beta}$ is fitted to the data found in the reference indicated.}
\begin{ruledtabular}
\begin{tabular}{c|llll}
compound & $\gamma$ ($\rm \frac{mJ}{mol \cdot K^2}$) & $P_c$ (GPa) & $T_N$ (K) & $T_{\beta}$ (K)   \tabularnewline
 \hline 
$ \rm CeIn_3$ &  130  & 2.65 \cite{Knebel_2001} & 10.23(1) \cite{Lawrence_1980} & 9.31(3) \tabularnewline
$ \rm CeRhIn_5$ &  70 \cite{Fisher_2002} & 2.3 \cite{ronning2017electronic} & 3.8 \cite{Hegger_2000} & 7.41(7) \cite{Cornelius_2001} \tabularnewline
$ \rm Ce_2RhIn_8$ &  400 \cite{Ueda_2004} & 1.36 \cite{Yashima_2010} & 2.8 \cite{Ueda_2004} & 4.445(14) \cite{Cornelius_2001} \tabularnewline
$ \rm CePt_2In_7$ &  50 \cite{Bauer_2010_pressure} & 3.5 \cite{Bauer_2010_pressure} & 5.5 \cite{Bauer_2010_pressure} & 7.02(7) \cite{Bauer_2010_pressure} \tabularnewline
$ \rm CePd_5Al_2$ &  56 \cite{Honda_2008_Pressure} & 10.8 \cite{Honda_2008_Pressure} & 2.87 \cite{Honda_2008_Pressure} & 3.274 \cite{Onimaru_2008} \tabularnewline
$ \rm CeCu_2Si_2$ &  1000 \cite{Steglich_1996} & 0 \cite{Steglich_1996} & 0 \cite{Steglich_1996} &  \tabularnewline
$ \rm CePd_2Si_2$ &  131 \cite{Sheikin_2002} & 2.87 \cite{Demuer_2001} & 9.3 \cite{Sheikin_2002} & 8.072(15) \cite{Sheikin_2002} \tabularnewline
$ \rm CeRh_2Si_2$ &  22.8 \cite{MOVSHOVICH1996126} & 0.97 \cite{Demuer_2001} & 36 \cite{QUEZEL1984685} & 56(5) \cite{Graf_1998} \tabularnewline
$ \rm CeCu_2Ge_2$ &  77 \cite{DEBOER198791} & 7.7 \cite{FISHER1994459} & 4.15(5) \cite{DEBOER198791} & 3.51(2) \cite{DEBOER198791} \tabularnewline
$ \rm Ce_2Ni_3Ge_5$ &  90 \cite{Thamizhavel_2005} & 3.9 \cite{Nakashima_2005} & 4.3 \cite{Thamizhavel_2005} & 3.551(7) \cite{Thamizhavel_2005} \tabularnewline
$ \rm CeNiGe_3$ &  76 \cite{Mun_2010_Tuning} & 5.5 \cite{Nakashima_2004} & 5 \cite{Mun_2010_Tuning} & 5.166(9) \cite{Mun_2010_Tuning} \tabularnewline
$ \rm CePt_3Si$ &  335 \cite{Takeuchi_2007} & 0.6 \cite{Takeuchi_2007} & 2.25 \cite{Takeuchi_2007} & 3.659(9) \cite{Takeuchi_2007} \tabularnewline
$ \rm CeRhSi_3$ &  110 \cite{Kimura_2007} & 2.36 \cite{Patztorova_2019} & 1.6 \cite{Kimura_2007} & 3.44(2) \cite{Kimura_2007} \tabularnewline
$ \rm CeIrSi_3$ &  105 \cite{Okuda_2007} & 2.63 \cite{Settai_2011} & 5 \cite{Okuda_2007} & 9.5(3) \cite{Okuda_2007} \tabularnewline
$ \rm CeCoGe_3$ &  32 \cite{Knebel_2009} & 5.5 \cite{Knebel_2009} & 21 \cite{Thamizhavel_2005_Unique} & 28.0(9) \cite{Thamizhavel_2005_Unique} \tabularnewline
$ \rm CePdAl$ &  250 \cite{Fritsch_2017} & 0.92 \cite{Majumder_2022} & 2.7 \cite{Fritsch_2017} & 2.862(7) \cite{Fritsch_2017} \tabularnewline
$ \rm CeRh_6Ge_4$ &  250 \cite{Matsuoka_2015} & 0.85 \cite{Kotegawa_2019} & 2.5 \cite{Matsuoka_2015} & 4.19(5) \cite{Matsuoka_2015} \tabularnewline
\end{tabular}\end{ruledtabular}
\label{tab:compounds}
\end{table}

Interestingly, we find many Ce-based heavy fermion materials with large low temperature $T^3$ heat capacity. In some cases this exists over a full decade in temperature. 
One might wonder if this is correlated with how ``close'' the system is to quantum criticality. If we take the critical pressure $P_c$ (at which magnetic ordering temperature goes to $T=0$) as a measure of this, we can answer this question empirically.

\begin{figure}
	\centering
	\includegraphics[width=0.42\textwidth]{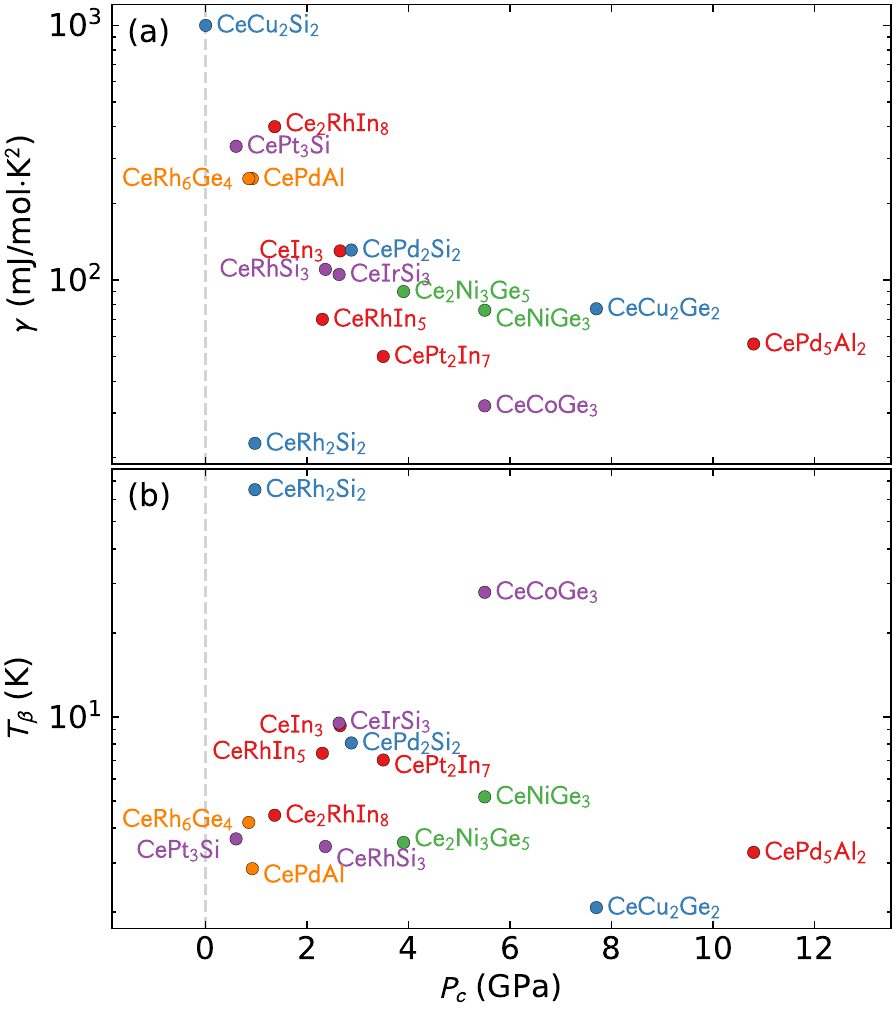}
	\caption{Correlations between physical properties for various magnetically ordered heavy fermion materials. (a) Sommerfeld coefficient $\gamma$ vs critical pressure $P_c$, showing a clear trend of increasing $\gamma$ as $P_c$ decreases. (b) $T^3$ term $\beta$ vs $P_c$, showing no clear correlation. Colors indicate families of materials. Data for this plot is shown in Table \ref{tab:compounds}.}
	\label{fig:CompoundsPc}
\end{figure}

Figure \ref{fig:CompoundsPc} plots the $\gamma$ and $T_{\beta}$ terms of the various compounds against $P_c$. For $\gamma$, there is a clear trend: the closer to criticality, the larger the $\gamma$  (with one outlier, CeRh$_2$Si$_2$ which has also an anomalously large $T_N$). This is as expected for mass renormalization driven by quantum criticality. For $T_{\beta}$ however, there is no apparent trend: the $T^3$ specific heat appears to be uncorrelated with $P_c$.

\begin{figure}
	\centering
	\includegraphics[width=0.44\textwidth]{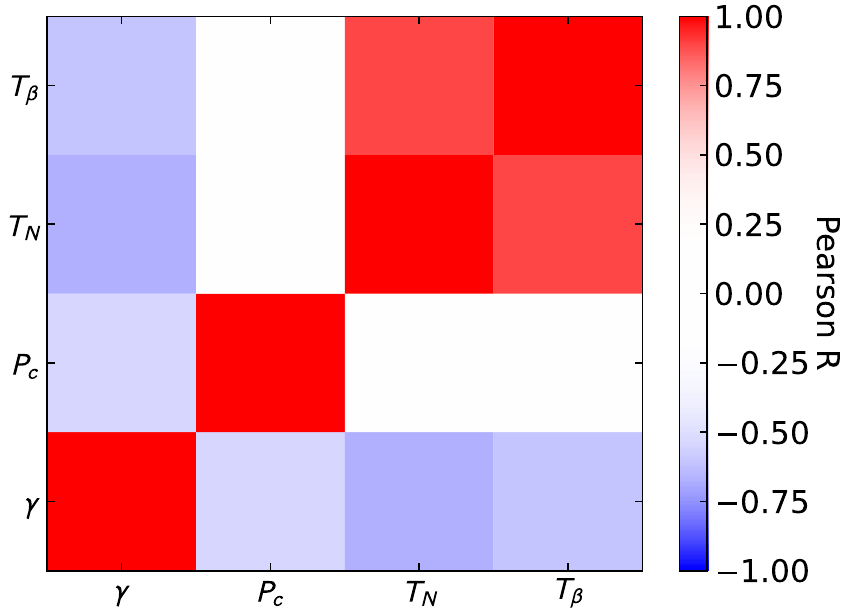}
	\caption{Correlation matrix between physical properties as indicated by the Pearson R coefficient. Red indicates positive correlation, blue indicates negative correlation. $T_{\beta}$ is strongly correlated only with $T_N$, and weakly correlated with $\gamma$, but essentially uncorrelated with $P_c$, indicating that it is not a function of proximity to a QCP.}
	\label{fig:CorrelationMatrix}
\end{figure}

We can be more precise about these trends by using Pearson's $R$ correlation coefficient. Applying this to the logarithm of the data (to account for nonlinear trends) in Table \ref{tab:compounds}  yields a correlation matrix, plotted in Fig. \ref{fig:CorrelationMatrix}. 
This reveals a very strong correlation between $T_{\beta}$ and $T_N$, weak correlation between $T_{\beta}$ and $\gamma$, and virtually no correlation between $T_{\beta}$ and $P_c$. 
Therefore, the $T^3$ heat capacity is not dependent on proximity to quantum criticality, but instead seems to be closely related to $T_N$ (indeed, in the Supplemental Materials \cite{SuppMat}, we show this relationship is essentially linear). Thus, this excess density of states seems to be governed by magnetic exchange interactions. 

This is consistent with behavior of CeRhIn$_5$ under hydrostatic pressure: as this compound approaches the QCP, the $\gamma$ value grows but $T_{\beta}$ shrinks as $T_N$ is suppressed \cite{Fisher_2002}. This was interpreted as ``decreasing spin wave stiffness,'' but our results here indicate that it is not spin waves at all, but of some other origin.  


One weakness of the above correlation analysis is that it does not consider how much of the $T^3$ heat capacity comes from magnons alone. However, the examples of CeIn$_3$ and other compounds in Fig. \ref{fig:HCmagnons} show the local-moment magnons come nowhere near explaining the heat capacity in the magnetically ordered state, suggesting it holds across the heavy fermion family. 

One additional example, not included in Fig. \ref{fig:HCmagnons}, is CeRh$_6$Ge$_4$. This material is a ferromagnet \cite{Matsuoka_2015}, which should have $T^{3/2}$ heat capacity at low temperatures because of its quadratic magnon dispersion \cite{AshcroftMermin}, but the magnetic specific heat is also definitively $T^3$ below $T_N$ (see the Supplemental Materials \cite{SuppMat}). This alone signals a significant discrepancy, but because there is no magnon model it is difficult to say how severe is the difference between the measured and magnon heat capacity. 

At this point, we are left with a quandary. We have shown that a large number of magnetically ordered Ce heavy fermion materials have an anomalous temperature-dependent heat capacity which often approximates $T^3$ as $T \rightarrow 0$, and this term is not related to QCP proximity. 
It is tempting to invoke heretofore unobserved Dirac fermions to explain these density of states. After all, a Dirac cone dispersion (linear dispersing bands) generically produces $T^3$ specific heat, and proposed Weyl-Kondo semimetal states in heavy fermions predict precisely such density of states at the Fermi energy \cite{Dzsaber_2017,lai2018weyl,Chen2022}. 
If this explanation is correct, it indicates that such behavior is far more common in the heavy fermions than previously thought. 
However, this explanation does not readily explain the correlation with $T_N$. Furthermore, because the velocity would have to be very small, it constrains the linear crossing to be close to the Fermi energy, for which no mechanism is known. 


Generically, coupling to bosonic fluctuations (e.g electron-hole pairs, magnons, or phonons) will also create a $T^3 \ln T$ contribution to the specific heat of a Fermi liquid \cite{ChubukovPRB2006}. However, such a Fermi liquid correction from electron-hole pairs can be ruled out because the correction is the wrong sign from that which is observed.  A correction due to coupling to magnons can also be ruled out on the basis that a magnetic field will gap out the magnons, while the CeIn$_3$  experimental heat capacity is essentially unchanged up to 9~T [Fig. \ref{fig:CeIn3}(c)] (similar low-field-independence is observed in the other  Fig. \ref{fig:HCmagnons} compounds \cite{Mishra_2021,Sheikin_2002,DEBOER198791,Takeuchi_2007}).  Finally, a correction from  coupling to phonons appears inconsistent with the observed correlation between $T_{\beta}$ and $T_N$, which suggests a magnetic origin; and the size of the excess heat capacity relative to the electronic term implies something beyond a perturbative correction to the Sommerfeld term. 
  

This observation of a pseudogapped density of states in magnetically ordered heavy fermions begs for an explanation. As it cannot be explained by electrons, magnons, or phonons alone, it suggests an entanglement between various degrees of freedom. 
For instance, it could be that the excess $T^3$ heat capacity arises from coherent spin waves in the itinerant electron bands that lie below the particle-hole continuum \cite{Platzman_1967}. If such physics could be produced by a staggered field of magnetic order (which remains to be seen), one could have density of states governed by magnetic exchange but in the itinerant electron bands---but this is speculation at this point. 

An interesting question, but beyond the scope of this study, is how common the pseudogap feature is in other types of compounds. A similar pseudogapped density of states has been observed in non-magnetically-ordered  Ce$_3$Bi$_4$Pd$_3$ \cite{Kushwaha2019} and elemental plutonium  \cite{Wartenbe_2022}, indicating generic heavy fermion behavior even beyond magnetically ordered systems. 
Another interesting question which may be addressed with a broader survey of compounds is whether the pseudogapped density of states correlates with the sharpness of the magnetic transition. 

In summary, we have shown that a large number of magnetically ordered Ce heavy fermions display anomalous, substantial $T^3$ specific heat inside their magnetic ordered phases. 
Comparison to the few materials for which magnon models exist shows that this heat capacity is not due to local-moment magnons. The $T^3$ term is not correlated with the critical pressure, indicating this effect is not due to QCP proximity; but is strongly correlated with the ordering temperature $T_N$, indicating the effect is governed by the magnetic exchange interaction.  
These results highlight a previously unobserved behavior: a Sommerfeld $\sim \gamma T$ term is insufficient to capture the density of states of the magnetically ordered Ce materials. 
Although it is perhaps not surprising that simplistic local-moment models fail to describe strongly correlated systems like magnetically ordered heavy fermions, this study highlights exactly how such models fail, and shows the fruitfulness of examining correlations across materials families. More importantly, the identified pseudogap will hopefully sharpen the theoretical studies of this fascinating class of strongly correlated materials. 

\section*{Acknowledgments}
\begin{acknowledgments} 
We gratefully acknowledge the U.S. Department of Energy, Office of Basic Energy Sciences, Division of Materials Science and Engineering under project ``Quantum Fluctuations in Narrow-Band Systems.'' DFT calculations were done with support from LANL LDRD Program.  
The authors also acknowledge helpful discussions with Shizeng Lin and Cristian Batista. 
\end{acknowledgments}


\begin{thebibliography}{70}%
	\makeatletter
	\providecommand \@ifxundefined [1]{%
		\@ifx{#1\undefined}
	}%
	\providecommand \@ifnum [1]{%
		\ifnum #1\expandafter \@firstoftwo
		\else \expandafter \@secondoftwo
		\fi
	}%
	\providecommand \@ifx [1]{%
		\ifx #1\expandafter \@firstoftwo
		\else \expandafter \@secondoftwo
		\fi
	}%
	\providecommand \natexlab [1]{#1}%
	\providecommand \enquote  [1]{``#1''}%
	\providecommand \bibnamefont  [1]{#1}%
	\providecommand \bibfnamefont [1]{#1}%
	\providecommand \citenamefont [1]{#1}%
	\providecommand \href@noop [0]{\@secondoftwo}%
	\providecommand \href [0]{\begingroup \@sanitize@url \@href}%
	\providecommand \@href[1]{\@@startlink{#1}\@@href}%
	\providecommand \@@href[1]{\endgroup#1\@@endlink}%
	\providecommand \@sanitize@url [0]{\catcode `\\12\catcode `\$12\catcode
		`\&12\catcode `\#12\catcode `\^12\catcode `\_12\catcode `\%12\relax}%
	\providecommand \@@startlink[1]{}%
	\providecommand \@@endlink[0]{}%
	\providecommand \url  [0]{\begingroup\@sanitize@url \@url }%
	\providecommand \@url [1]{\endgroup\@href {#1}{\urlprefix }}%
	\providecommand \urlprefix  [0]{URL }%
	\providecommand \Eprint [0]{\href }%
	\providecommand \doibase [0]{https://doi.org/}%
	\providecommand \selectlanguage [0]{\@gobble}%
	\providecommand \bibinfo  [0]{\@secondoftwo}%
	\providecommand \bibfield  [0]{\@secondoftwo}%
	\providecommand \translation [1]{[#1]}%
	\providecommand \BibitemOpen [0]{}%
	\providecommand \bibitemStop [0]{}%
	\providecommand \bibitemNoStop [0]{.\EOS\space}%
	\providecommand \EOS [0]{\spacefactor3000\relax}%
	\providecommand \BibitemShut  [1]{\csname bibitem#1\endcsname}%
	\let\auto@bib@innerbib\@empty
	\bibitem [{\citenamefont {Andres}\ \emph {et~al.}(1975)\citenamefont {Andres},
		\citenamefont {Graebner},\ and\ \citenamefont {Ott}}]{Andres_1975}%
	\BibitemOpen
	\bibfield  {author} {\bibinfo {author} {\bibfnamefont {K.}~\bibnamefont
			{Andres}}, \bibinfo {author} {\bibfnamefont {J.~E.}\ \bibnamefont
			{Graebner}},\ and\ \bibinfo {author} {\bibfnamefont {H.~R.}\ \bibnamefont
			{Ott}},\ }\href {https://doi.org/10.1103/PhysRevLett.35.1779} {\bibfield
		{journal} {\bibinfo  {journal} {Phys. Rev. Lett.}\ }\textbf {\bibinfo
			{volume} {35}},\ \bibinfo {pages} {1779} (\bibinfo {year}
		{1975})}\BibitemShut {NoStop}%
	\bibitem [{\citenamefont {Steglich}\ \emph {et~al.}(1979)\citenamefont
		{Steglich}, \citenamefont {Aarts}, \citenamefont {Bredl}, \citenamefont
		{Lieke}, \citenamefont {Meschede}, \citenamefont {Franz},\ and\ \citenamefont
		{Sch\"afer}}]{Steglich_1979}%
	\BibitemOpen
	\bibfield  {author} {\bibinfo {author} {\bibfnamefont {F.}~\bibnamefont
			{Steglich}}, \bibinfo {author} {\bibfnamefont {J.}~\bibnamefont {Aarts}},
		\bibinfo {author} {\bibfnamefont {C.~D.}\ \bibnamefont {Bredl}}, \bibinfo
		{author} {\bibfnamefont {W.}~\bibnamefont {Lieke}}, \bibinfo {author}
		{\bibfnamefont {D.}~\bibnamefont {Meschede}}, \bibinfo {author}
		{\bibfnamefont {W.}~\bibnamefont {Franz}},\ and\ \bibinfo {author}
		{\bibfnamefont {H.}~\bibnamefont {Sch\"afer}},\ }\href
	{https://doi.org/10.1103/PhysRevLett.43.1892} {\bibfield  {journal} {\bibinfo
			{journal} {Phys. Rev. Lett.}\ }\textbf {\bibinfo {volume} {43}},\ \bibinfo
		{pages} {1892} (\bibinfo {year} {1979})}\BibitemShut {NoStop}%
	\bibitem [{\citenamefont {Stewart}(1984)}]{Steward_1984_RMP}%
	\BibitemOpen
	\bibfield  {author} {\bibinfo {author} {\bibfnamefont {G.~R.}\ \bibnamefont
			{Stewart}},\ }\href {https://doi.org/10.1103/RevModPhys.56.755} {\bibfield
		{journal} {\bibinfo  {journal} {Rev. Mod. Phys.}\ }\textbf {\bibinfo {volume}
			{56}},\ \bibinfo {pages} {755} (\bibinfo {year} {1984})}\BibitemShut
	{NoStop}%
	\bibitem [{\citenamefont {Fisk}\ \emph {et~al.}(1995)\citenamefont {Fisk},
		\citenamefont {Sarrao}, \citenamefont {Smith},\ and\ \citenamefont
		{Thompson}}]{fisk1995physics}%
	\BibitemOpen
	\bibfield  {author} {\bibinfo {author} {\bibfnamefont {Z.}~\bibnamefont
			{Fisk}}, \bibinfo {author} {\bibfnamefont {J.}~\bibnamefont {Sarrao}},
		\bibinfo {author} {\bibfnamefont {J.}~\bibnamefont {Smith}},\ and\ \bibinfo
		{author} {\bibfnamefont {J.}~\bibnamefont {Thompson}},\ }\href
	{https://doi.org/10.1073/pnas.92.15.6663} {\bibfield  {journal} {\bibinfo
			{journal} {Proceedings of the National Academy of Sciences}\ }\textbf
		{\bibinfo {volume} {92}},\ \bibinfo {pages} {6663} (\bibinfo {year}
		{1995})}\BibitemShut {NoStop}%
	\bibitem [{\citenamefont {Pavarini}\ \emph {et~al.}(2015)\citenamefont
		{Pavarini}, \citenamefont {Coleman},\ and\ \citenamefont
		{Koch}}]{pavarini2015many}%
	\BibitemOpen
	\bibfield  {author} {\bibinfo {author} {\bibfnamefont {E.}~\bibnamefont
			{Pavarini}}, \bibinfo {author} {\bibfnamefont {P.}~\bibnamefont {Coleman}},\
		and\ \bibinfo {author} {\bibfnamefont {E.}~\bibnamefont {Koch}},\ }\href
	{http://hdl.handle.net/2128/9255} {\emph {\bibinfo {title} {Many-body
				physics: from Kondo to Hubbard}}},\ \bibinfo {type} {Tech. Rep.}\ (\bibinfo
	{institution} {Theoretische Nanoelektronik},\ \bibinfo {year}
	{2015})\BibitemShut {NoStop}%
	\bibitem [{\citenamefont {Wirth}\ and\ \citenamefont
		{Steglich}(2016)}]{Wirth2016}%
	\BibitemOpen
	\bibfield  {author} {\bibinfo {author} {\bibfnamefont {S.}~\bibnamefont
			{Wirth}}\ and\ \bibinfo {author} {\bibfnamefont {F.}~\bibnamefont
			{Steglich}},\ }\href {https://doi.org/10.1038/natrevmats.2016.51} {\bibfield
		{journal} {\bibinfo  {journal} {Nature Reviews Materials}\ }\textbf {\bibinfo
			{volume} {1}},\ \bibinfo {pages} {16051} (\bibinfo {year}
		{2016})}\BibitemShut {NoStop}%
	\bibitem [{\citenamefont {Steglich}\ and\ \citenamefont
		{Wirth}(2016)}]{Steglich_2016}%
	\BibitemOpen
	\bibfield  {author} {\bibinfo {author} {\bibfnamefont {F.}~\bibnamefont
			{Steglich}}\ and\ \bibinfo {author} {\bibfnamefont {S.}~\bibnamefont
			{Wirth}},\ }\href {https://doi.org/10.1088/0034-4885/79/8/084502} {\bibfield
		{journal} {\bibinfo  {journal} {Reports on Progress in Physics}\ }\textbf
		{\bibinfo {volume} {79}},\ \bibinfo {pages} {084502} (\bibinfo {year}
		{2016})}\BibitemShut {NoStop}%
	\bibitem [{\citenamefont {Shaginyan}\ \emph {et~al.}(2022)\citenamefont
		{Shaginyan}, \citenamefont {Msezane},\ and\ \citenamefont
		{Japaridze}}]{shaginyan2022peculiar}%
	\BibitemOpen
	\bibfield  {author} {\bibinfo {author} {\bibfnamefont {V.~R.}\ \bibnamefont
			{Shaginyan}}, \bibinfo {author} {\bibfnamefont {A.~Z.}\ \bibnamefont
			{Msezane}},\ and\ \bibinfo {author} {\bibfnamefont {G.~S.}\ \bibnamefont
			{Japaridze}},\ }\href {https://doi.org/10.3390/atoms10030067} {\bibfield
		{journal} {\bibinfo  {journal} {Atoms}\ }\textbf {\bibinfo {volume} {10}},\
		\bibinfo {pages} {67} (\bibinfo {year} {2022})}\BibitemShut {NoStop}%
	\bibitem [{\citenamefont {Si}\ and\ \citenamefont
		{Steglich}(2010)}]{si2010heavy}%
	\BibitemOpen
	\bibfield  {author} {\bibinfo {author} {\bibfnamefont {Q.}~\bibnamefont
			{Si}}\ and\ \bibinfo {author} {\bibfnamefont {F.}~\bibnamefont {Steglich}},\
	}\href {https://doi.org/10.1126/science.1191195} {\bibfield  {journal}
		{\bibinfo  {journal} {Science}\ }\textbf {\bibinfo {volume} {329}},\ \bibinfo
		{pages} {1161} (\bibinfo {year} {2010})}\BibitemShut {NoStop}%
	\bibitem [{\citenamefont {Lavagna}\ \emph {et~al.}(1982)\citenamefont
		{Lavagna}, \citenamefont {Lacroix},\ and\ \citenamefont
		{Cyrot}}]{LAVAGNA1982210}%
	\BibitemOpen
	\bibfield  {author} {\bibinfo {author} {\bibfnamefont {M.}~\bibnamefont
			{Lavagna}}, \bibinfo {author} {\bibfnamefont {C.}~\bibnamefont {Lacroix}},\
		and\ \bibinfo {author} {\bibfnamefont {M.}~\bibnamefont {Cyrot}},\ }\href
	{https://doi.org/https://doi.org/10.1016/0375-9601(82)90689-2} {\bibfield
		{journal} {\bibinfo  {journal} {Physics Letters A}\ }\textbf {\bibinfo
			{volume} {90}},\ \bibinfo {pages} {210} (\bibinfo {year} {1982})}\BibitemShut
	{NoStop}%
	\bibitem [{\citenamefont {Dzero}\ \emph {et~al.}(2010)\citenamefont {Dzero},
		\citenamefont {Sun}, \citenamefont {Galitski},\ and\ \citenamefont
		{Coleman}}]{Dzero_2010}%
	\BibitemOpen
	\bibfield  {author} {\bibinfo {author} {\bibfnamefont {M.}~\bibnamefont
			{Dzero}}, \bibinfo {author} {\bibfnamefont {K.}~\bibnamefont {Sun}}, \bibinfo
		{author} {\bibfnamefont {V.}~\bibnamefont {Galitski}},\ and\ \bibinfo
		{author} {\bibfnamefont {P.}~\bibnamefont {Coleman}},\ }\href
	{https://doi.org/10.1103/PhysRevLett.104.106408} {\bibfield  {journal}
		{\bibinfo  {journal} {Phys. Rev. Lett.}\ }\textbf {\bibinfo {volume} {104}},\
		\bibinfo {pages} {106408} (\bibinfo {year} {2010})}\BibitemShut {NoStop}%
	\bibitem [{\citenamefont {Mydosh}\ and\ \citenamefont
		{Oppeneer}(2014)}]{Mydosh_2014}%
	\BibitemOpen
	\bibfield  {author} {\bibinfo {author} {\bibfnamefont {J.}~\bibnamefont
			{Mydosh}}\ and\ \bibinfo {author} {\bibfnamefont {P.}~\bibnamefont
			{Oppeneer}},\ }\href {https://doi.org/10.1080/14786435.2014.916428}
	{\bibfield  {journal} {\bibinfo  {journal} {Philosophical Magazine}\ }\textbf
		{\bibinfo {volume} {94}},\ \bibinfo {pages} {3642} (\bibinfo {year}
		{2014})}\BibitemShut {NoStop}%
	\bibitem [{\citenamefont {Doniach}(1977)}]{Doniach1977}%
	\BibitemOpen
	\bibfield  {author} {\bibinfo {author} {\bibfnamefont {S.}~\bibnamefont
			{Doniach}},\ }\bibinfo {title} {Phase diagram for the kondo lattice},\ in\
	\href {https://doi.org/10.1007/978-1-4615-8816-0_15} {\emph {\bibinfo
			{booktitle} {Valence Instabilities and Related Narrow-Band Phenomena}}},\
	\bibinfo {editor} {edited by\ \bibinfo {editor} {\bibfnamefont {R.~D.}\
			\bibnamefont {Parks}}}\ (\bibinfo  {publisher} {Springer US},\ \bibinfo
	{address} {Boston, MA},\ \bibinfo {year} {1977})\ pp.\ \bibinfo {pages}
	{169--176}\BibitemShut {NoStop}%
	\bibitem [{\citenamefont {Kadowaki}\ and\ \citenamefont
		{Woods}(1986)}]{KADOWAKI1986507}%
	\BibitemOpen
	\bibfield  {author} {\bibinfo {author} {\bibfnamefont {K.}~\bibnamefont
			{Kadowaki}}\ and\ \bibinfo {author} {\bibfnamefont {S.}~\bibnamefont
			{Woods}},\ }\href
	{https://doi.org/https://doi.org/10.1016/0038-1098(86)90785-4} {\bibfield
		{journal} {\bibinfo  {journal} {Solid State Communications}\ }\textbf
		{\bibinfo {volume} {58}},\ \bibinfo {pages} {507} (\bibinfo {year}
		{1986})}\BibitemShut {NoStop}%
	\bibitem [{\citenamefont {Wilson}(1975)}]{Wilson_1975}%
	\BibitemOpen
	\bibfield  {author} {\bibinfo {author} {\bibfnamefont {K.~G.}\ \bibnamefont
			{Wilson}},\ }\href {https://doi.org/10.1103/RevModPhys.47.773} {\bibfield
		{journal} {\bibinfo  {journal} {Rev. Mod. Phys.}\ }\textbf {\bibinfo {volume}
			{47}},\ \bibinfo {pages} {773} (\bibinfo {year} {1975})}\BibitemShut
	{NoStop}%
	\bibitem [{\citenamefont {Weng}\ \emph {et~al.}(2016)\citenamefont {Weng},
		\citenamefont {Smidman}, \citenamefont {Jiao}, \citenamefont {Lu},\ and\
		\citenamefont {Yuan}}]{Weng_2016}%
	\BibitemOpen
	\bibfield  {author} {\bibinfo {author} {\bibfnamefont {Z.~F.}\ \bibnamefont
			{Weng}}, \bibinfo {author} {\bibfnamefont {M.}~\bibnamefont {Smidman}},
		\bibinfo {author} {\bibfnamefont {L.}~\bibnamefont {Jiao}}, \bibinfo {author}
		{\bibfnamefont {X.}~\bibnamefont {Lu}},\ and\ \bibinfo {author}
		{\bibfnamefont {H.~Q.}\ \bibnamefont {Yuan}},\ }\href
	{https://doi.org/10.1088/0034-4885/79/9/094503} {\bibfield  {journal}
		{\bibinfo  {journal} {Reports on Progress in Physics}\ }\textbf {\bibinfo
			{volume} {79}},\ \bibinfo {pages} {094503} (\bibinfo {year}
		{2016})}\BibitemShut {NoStop}%
	\bibitem [{\citenamefont {Lawrence}\ and\ \citenamefont
		{Shapiro}(1980)}]{Lawrence_1980}%
	\BibitemOpen
	\bibfield  {author} {\bibinfo {author} {\bibfnamefont {J.~M.}\ \bibnamefont
			{Lawrence}}\ and\ \bibinfo {author} {\bibfnamefont {S.~M.}\ \bibnamefont
			{Shapiro}},\ }\href {https://doi.org/10.1103/PhysRevB.22.4379} {\bibfield
		{journal} {\bibinfo  {journal} {Phys. Rev. B}\ }\textbf {\bibinfo {volume}
			{22}},\ \bibinfo {pages} {4379} (\bibinfo {year} {1980})}\BibitemShut
	{NoStop}%
	\bibitem [{\citenamefont {Mathur}\ \emph {et~al.}(1998)\citenamefont {Mathur},
		\citenamefont {Grosche}, \citenamefont {Julian}, \citenamefont {Walker},
		\citenamefont {Freye}, \citenamefont {Haselwimmer},\ and\ \citenamefont
		{Lonzarich}}]{Mathur1998Nature}%
	\BibitemOpen
	\bibfield  {author} {\bibinfo {author} {\bibfnamefont {N.~D.}\ \bibnamefont
			{Mathur}}, \bibinfo {author} {\bibfnamefont {F.~M.}\ \bibnamefont {Grosche}},
		\bibinfo {author} {\bibfnamefont {S.~R.}\ \bibnamefont {Julian}}, \bibinfo
		{author} {\bibfnamefont {I.~R.}\ \bibnamefont {Walker}}, \bibinfo {author}
		{\bibfnamefont {D.~M.}\ \bibnamefont {Freye}}, \bibinfo {author}
		{\bibfnamefont {R.~K.~W.}\ \bibnamefont {Haselwimmer}},\ and\ \bibinfo
		{author} {\bibfnamefont {G.~G.}\ \bibnamefont {Lonzarich}},\ }\href
	{https://doi.org/10.1038/27838} {\bibfield  {journal} {\bibinfo  {journal}
			{Nature}\ }\textbf {\bibinfo {volume} {394}},\ \bibinfo {pages} {39}
		(\bibinfo {year} {1998})}\BibitemShut {NoStop}%
	\bibitem [{\citenamefont {Knebel}\ \emph {et~al.}(2001)\citenamefont {Knebel},
		\citenamefont {Braithwaite}, \citenamefont {Canfield}, \citenamefont
		{Lapertot},\ and\ \citenamefont {Flouquet}}]{Knebel_2001}%
	\BibitemOpen
	\bibfield  {author} {\bibinfo {author} {\bibfnamefont {G.}~\bibnamefont
			{Knebel}}, \bibinfo {author} {\bibfnamefont {D.}~\bibnamefont {Braithwaite}},
		\bibinfo {author} {\bibfnamefont {P.~C.}\ \bibnamefont {Canfield}}, \bibinfo
		{author} {\bibfnamefont {G.}~\bibnamefont {Lapertot}},\ and\ \bibinfo
		{author} {\bibfnamefont {J.}~\bibnamefont {Flouquet}},\ }\href
	{https://doi.org/10.1103/PhysRevB.65.024425} {\bibfield  {journal} {\bibinfo
			{journal} {Phys. Rev. B}\ }\textbf {\bibinfo {volume} {65}},\ \bibinfo
		{pages} {024425} (\bibinfo {year} {2001})}\BibitemShut {NoStop}%
	\bibitem [{\citenamefont {Ashcroft}\ and\ \citenamefont
		{Mermin}(1976)}]{AshcroftMermin}%
	\BibitemOpen
	\bibfield  {author} {\bibinfo {author} {\bibfnamefont {N.~W.}\ \bibnamefont
			{Ashcroft}}\ and\ \bibinfo {author} {\bibfnamefont {N.~D.}\ \bibnamefont
			{Mermin}},\ }\href@noop {} {\emph {\bibinfo {title} {Solid state physics}}}\
	(\bibinfo  {publisher} {Cengage Learning},\ \bibinfo {year}
	{1976})\BibitemShut {NoStop}%
	\bibitem [{\citenamefont {Berry}\ \emph {et~al.}(2010)\citenamefont {Berry},
		\citenamefont {Bittar}, \citenamefont {Capan}, \citenamefont {Pagliuso},\
		and\ \citenamefont {Fisk}}]{Berry_2010}%
	\BibitemOpen
	\bibfield  {author} {\bibinfo {author} {\bibfnamefont {N.}~\bibnamefont
			{Berry}}, \bibinfo {author} {\bibfnamefont {E.~M.}\ \bibnamefont {Bittar}},
		\bibinfo {author} {\bibfnamefont {C.}~\bibnamefont {Capan}}, \bibinfo
		{author} {\bibfnamefont {P.~G.}\ \bibnamefont {Pagliuso}},\ and\ \bibinfo
		{author} {\bibfnamefont {Z.}~\bibnamefont {Fisk}},\ }\href
	{https://doi.org/10.1103/PhysRevB.81.174413} {\bibfield  {journal} {\bibinfo
			{journal} {Phys. Rev. B}\ }\textbf {\bibinfo {volume} {81}},\ \bibinfo
		{pages} {174413} (\bibinfo {year} {2010})}\BibitemShut {NoStop}%
	\bibitem [{\citenamefont {Simeth}\ \emph {et~al.}(2023)\citenamefont {Simeth},
		\citenamefont {Wang}, \citenamefont {Ghioldi}, \citenamefont {Fobes},
		\citenamefont {Podlesnyak}, \citenamefont {Sung}, \citenamefont {Bauer},
		\citenamefont {Lass}, \citenamefont {Flury}, \citenamefont {Vonka},
		\citenamefont {Mazzone}, \citenamefont {Niedermayer}, \citenamefont {Nomura},
		\citenamefont {Arita}, \citenamefont {Batista}, \citenamefont {Ronning},\
		and\ \citenamefont {Janoschek}}]{simeth2022microscopic}%
	\BibitemOpen
	\bibfield  {author} {\bibinfo {author} {\bibfnamefont {W.}~\bibnamefont
			{Simeth}}, \bibinfo {author} {\bibfnamefont {Z.}~\bibnamefont {Wang}},
		\bibinfo {author} {\bibfnamefont {E.~A.}\ \bibnamefont {Ghioldi}}, \bibinfo
		{author} {\bibfnamefont {D.~M.}\ \bibnamefont {Fobes}}, \bibinfo {author}
		{\bibfnamefont {A.}~\bibnamefont {Podlesnyak}}, \bibinfo {author}
		{\bibfnamefont {N.~H.}\ \bibnamefont {Sung}}, \bibinfo {author}
		{\bibfnamefont {E.~D.}\ \bibnamefont {Bauer}}, \bibinfo {author}
		{\bibfnamefont {J.}~\bibnamefont {Lass}}, \bibinfo {author} {\bibfnamefont
			{S.}~\bibnamefont {Flury}}, \bibinfo {author} {\bibfnamefont
			{J.}~\bibnamefont {Vonka}}, \bibinfo {author} {\bibfnamefont {D.~G.}\
			\bibnamefont {Mazzone}}, \bibinfo {author} {\bibfnamefont {C.}~\bibnamefont
			{Niedermayer}}, \bibinfo {author} {\bibfnamefont {Y.}~\bibnamefont {Nomura}},
		\bibinfo {author} {\bibfnamefont {R.}~\bibnamefont {Arita}}, \bibinfo
		{author} {\bibfnamefont {C.~D.}\ \bibnamefont {Batista}}, \bibinfo {author}
		{\bibfnamefont {F.}~\bibnamefont {Ronning}},\ and\ \bibinfo {author}
		{\bibfnamefont {M.}~\bibnamefont {Janoschek}},\ }\href
	{https://doi.org/10.1038/s41467-023-43947-z} {\bibfield  {journal} {\bibinfo
			{journal} {Nature Communications}\ }\textbf {\bibinfo {volume} {14}},\
		\bibinfo {pages} {8239} (\bibinfo {year} {2023})}\BibitemShut {NoStop}%
	\bibitem [{\citenamefont {Blaha}\ \emph {et~al.}(2001)\citenamefont {Blaha},
		\citenamefont {Schwarz}, \citenamefont {Madsen}, \citenamefont {Kvasnicka},
		\citenamefont {Luitz} \emph {et~al.}}]{PBlaha:2001}%
	\BibitemOpen
	\bibfield  {author} {\bibinfo {author} {\bibfnamefont {P.}~\bibnamefont
			{Blaha}}, \bibinfo {author} {\bibfnamefont {K.}~\bibnamefont {Schwarz}},
		\bibinfo {author} {\bibfnamefont {G.~K.}\ \bibnamefont {Madsen}}, \bibinfo
		{author} {\bibfnamefont {D.}~\bibnamefont {Kvasnicka}}, \bibinfo {author}
		{\bibfnamefont {J.}~\bibnamefont {Luitz}}, \emph {et~al.},\ }\href@noop {}
	{\bibfield  {journal} {\bibinfo  {journal} {An augmented plane wave+ local
				orbitals program for calculating crystal properties}\ }\textbf {\bibinfo
			{volume} {60}} (\bibinfo {year} {2001})}\BibitemShut {NoStop}%
	\bibitem [{\citenamefont {Perdew}\ \emph {et~al.}(1996)\citenamefont {Perdew},
		\citenamefont {Burke},\ and\ \citenamefont {Ernzerhof}}]{Perdew_1996}%
	\BibitemOpen
	\bibfield  {author} {\bibinfo {author} {\bibfnamefont {J.~P.}\ \bibnamefont
			{Perdew}}, \bibinfo {author} {\bibfnamefont {K.}~\bibnamefont {Burke}},\ and\
		\bibinfo {author} {\bibfnamefont {M.}~\bibnamefont {Ernzerhof}},\ }\href
	{https://doi.org/10.1103/PhysRevLett.77.3865} {\bibfield  {journal} {\bibinfo
			{journal} {Phys. Rev. Lett.}\ }\textbf {\bibinfo {volume} {77}},\ \bibinfo
		{pages} {3865} (\bibinfo {year} {1996})}\BibitemShut {NoStop}%
	\bibitem [{\citenamefont {Desgranges}\ and\ \citenamefont
		{Schotte}(1982)}]{DESGRANGES1982}%
	\BibitemOpen
	\bibfield  {author} {\bibinfo {author} {\bibfnamefont {H.-U.}\ \bibnamefont
			{Desgranges}}\ and\ \bibinfo {author} {\bibfnamefont {K.}~\bibnamefont
			{Schotte}},\ }\href
	{https://doi.org/https://doi.org/10.1016/0375-9601(82)90481-9} {\bibfield
		{journal} {\bibinfo  {journal} {Physics Letters A}\ }\textbf {\bibinfo
			{volume} {91}},\ \bibinfo {pages} {240} (\bibinfo {year} {1982})}\BibitemShut
	{NoStop}%
	\bibitem [{\citenamefont {Das}\ \emph {et~al.}(2014)\citenamefont {Das},
		\citenamefont {Lin}, \citenamefont {Ghimire}, \citenamefont {Huang},
		\citenamefont {Ronning}, \citenamefont {Bauer}, \citenamefont {Thompson},
		\citenamefont {Batista}, \citenamefont {Ehlers},\ and\ \citenamefont
		{Janoschek}}]{Das_2014_CeRhIn5}%
	\BibitemOpen
	\bibfield  {author} {\bibinfo {author} {\bibfnamefont {P.}~\bibnamefont
			{Das}}, \bibinfo {author} {\bibfnamefont {S.-Z.}\ \bibnamefont {Lin}},
		\bibinfo {author} {\bibfnamefont {N.~J.}\ \bibnamefont {Ghimire}}, \bibinfo
		{author} {\bibfnamefont {K.}~\bibnamefont {Huang}}, \bibinfo {author}
		{\bibfnamefont {F.}~\bibnamefont {Ronning}}, \bibinfo {author} {\bibfnamefont
			{E.~D.}\ \bibnamefont {Bauer}}, \bibinfo {author} {\bibfnamefont {J.~D.}\
			\bibnamefont {Thompson}}, \bibinfo {author} {\bibfnamefont {C.~D.}\
			\bibnamefont {Batista}}, \bibinfo {author} {\bibfnamefont {G.}~\bibnamefont
			{Ehlers}},\ and\ \bibinfo {author} {\bibfnamefont {M.}~\bibnamefont
			{Janoschek}},\ }\href {https://doi.org/10.1103/PhysRevLett.113.246403}
	{\bibfield  {journal} {\bibinfo  {journal} {Phys. Rev. Lett.}\ }\textbf
		{\bibinfo {volume} {113}},\ \bibinfo {pages} {246403} (\bibinfo {year}
		{2014})}\BibitemShut {NoStop}%
	\bibitem [{\citenamefont {van Dijk}\ \emph {et~al.}(2000)\citenamefont {van
			Dijk}, \citenamefont {F\aa{}k}, \citenamefont {Charvolin}, \citenamefont
		{Lejay},\ and\ \citenamefont {Mignot}}]{Dijk_2000}%
	\BibitemOpen
	\bibfield  {author} {\bibinfo {author} {\bibfnamefont {N.~H.}\ \bibnamefont
			{van Dijk}}, \bibinfo {author} {\bibfnamefont {B.}~\bibnamefont {F\aa{}k}},
		\bibinfo {author} {\bibfnamefont {T.}~\bibnamefont {Charvolin}}, \bibinfo
		{author} {\bibfnamefont {P.}~\bibnamefont {Lejay}},\ and\ \bibinfo {author}
		{\bibfnamefont {J.~M.}\ \bibnamefont {Mignot}},\ }\href
	{https://doi.org/10.1103/PhysRevB.61.8922} {\bibfield  {journal} {\bibinfo
			{journal} {Phys. Rev. B}\ }\textbf {\bibinfo {volume} {61}},\ \bibinfo
		{pages} {8922} (\bibinfo {year} {2000})}\BibitemShut {NoStop}%
	\bibitem [{\citenamefont {Knopp}\ \emph {et~al.}(1989)\citenamefont {Knopp},
		\citenamefont {Loidl}, \citenamefont {Knorr}, \citenamefont {Pawlak},
		\citenamefont {Duczmal}, \citenamefont {Caspary}, \citenamefont {Gottwick},
		\citenamefont {Spille}, \citenamefont {Steglich},\ and\ \citenamefont
		{Murani}}]{Knopp1989}%
	\BibitemOpen
	\bibfield  {author} {\bibinfo {author} {\bibfnamefont {G.}~\bibnamefont
			{Knopp}}, \bibinfo {author} {\bibfnamefont {A.}~\bibnamefont {Loidl}},
		\bibinfo {author} {\bibfnamefont {K.}~\bibnamefont {Knorr}}, \bibinfo
		{author} {\bibfnamefont {L.}~\bibnamefont {Pawlak}}, \bibinfo {author}
		{\bibfnamefont {M.}~\bibnamefont {Duczmal}}, \bibinfo {author} {\bibfnamefont
			{R.}~\bibnamefont {Caspary}}, \bibinfo {author} {\bibfnamefont
			{U.}~\bibnamefont {Gottwick}}, \bibinfo {author} {\bibfnamefont
			{H.}~\bibnamefont {Spille}}, \bibinfo {author} {\bibfnamefont
			{F.}~\bibnamefont {Steglich}},\ and\ \bibinfo {author} {\bibfnamefont
			{A.~P.}\ \bibnamefont {Murani}},\ }\href {https://doi.org/10.1007/BF01313625}
	{\bibfield  {journal} {\bibinfo  {journal} {Zeitschrift f{\"u}r Physik B
				Condensed Matter}\ }\textbf {\bibinfo {volume} {77}},\ \bibinfo {pages} {95}
		(\bibinfo {year} {1989})}\BibitemShut {NoStop}%
	\bibitem [{\citenamefont {F\aa{}k}\ \emph {et~al.}(2008)\citenamefont
		{F\aa{}k}, \citenamefont {Raymond}, \citenamefont {Braithwaite},
		\citenamefont {Lapertot},\ and\ \citenamefont {Mignot}}]{Faak_2008}%
	\BibitemOpen
	\bibfield  {author} {\bibinfo {author} {\bibfnamefont {B.}~\bibnamefont
			{F\aa{}k}}, \bibinfo {author} {\bibfnamefont {S.}~\bibnamefont {Raymond}},
		\bibinfo {author} {\bibfnamefont {D.}~\bibnamefont {Braithwaite}}, \bibinfo
		{author} {\bibfnamefont {G.}~\bibnamefont {Lapertot}},\ and\ \bibinfo
		{author} {\bibfnamefont {J.-M.}\ \bibnamefont {Mignot}},\ }\href
	{https://doi.org/10.1103/PhysRevB.78.184518} {\bibfield  {journal} {\bibinfo
			{journal} {Phys. Rev. B}\ }\textbf {\bibinfo {volume} {78}},\ \bibinfo
		{pages} {184518} (\bibinfo {year} {2008})}\BibitemShut {NoStop}%
	\bibitem [{Sup()}]{SuppMat}%
	\BibitemOpen
	\href@noop {} {}\bibinfo {note} {See Supplemental Material at [URL will be
		inserted by publisher] for more details of the experiments and
		calculations.}\BibitemShut {Stop}%
	\bibitem [{\citenamefont {Fisher}\ \emph {et~al.}(2002)\citenamefont {Fisher},
		\citenamefont {Bouquet}, \citenamefont {Phillips}, \citenamefont {Hundley},
		\citenamefont {Pagliuso}, \citenamefont {Sarrao}, \citenamefont {Fisk},\ and\
		\citenamefont {Thompson}}]{Fisher_2002}%
	\BibitemOpen
	\bibfield  {author} {\bibinfo {author} {\bibfnamefont {R.~A.}\ \bibnamefont
			{Fisher}}, \bibinfo {author} {\bibfnamefont {F.}~\bibnamefont {Bouquet}},
		\bibinfo {author} {\bibfnamefont {N.~E.}\ \bibnamefont {Phillips}}, \bibinfo
		{author} {\bibfnamefont {M.~F.}\ \bibnamefont {Hundley}}, \bibinfo {author}
		{\bibfnamefont {P.~G.}\ \bibnamefont {Pagliuso}}, \bibinfo {author}
		{\bibfnamefont {J.~L.}\ \bibnamefont {Sarrao}}, \bibinfo {author}
		{\bibfnamefont {Z.}~\bibnamefont {Fisk}},\ and\ \bibinfo {author}
		{\bibfnamefont {J.~D.}\ \bibnamefont {Thompson}},\ }\href
	{https://doi.org/10.1103/PhysRevB.65.224509} {\bibfield  {journal} {\bibinfo
			{journal} {Phys. Rev. B}\ }\textbf {\bibinfo {volume} {65}},\ \bibinfo
		{pages} {224509} (\bibinfo {year} {2002})}\BibitemShut {NoStop}%
	\bibitem [{\citenamefont {Ronning}\ \emph {et~al.}(2017)\citenamefont
		{Ronning}, \citenamefont {Helm}, \citenamefont {Shirer}, \citenamefont
		{Bachmann}, \citenamefont {Balicas}, \citenamefont {Chan}, \citenamefont
		{Ramshaw}, \citenamefont {Mcdonald}, \citenamefont {Balakirev}, \citenamefont
		{Jaime} \emph {et~al.}}]{ronning2017electronic}%
	\BibitemOpen
	\bibfield  {author} {\bibinfo {author} {\bibfnamefont {F.}~\bibnamefont
			{Ronning}}, \bibinfo {author} {\bibfnamefont {T.}~\bibnamefont {Helm}},
		\bibinfo {author} {\bibfnamefont {K.}~\bibnamefont {Shirer}}, \bibinfo
		{author} {\bibfnamefont {M.}~\bibnamefont {Bachmann}}, \bibinfo {author}
		{\bibfnamefont {L.}~\bibnamefont {Balicas}}, \bibinfo {author} {\bibfnamefont
			{M.~K.}\ \bibnamefont {Chan}}, \bibinfo {author} {\bibfnamefont
			{B.}~\bibnamefont {Ramshaw}}, \bibinfo {author} {\bibfnamefont {R.~D.}\
			\bibnamefont {Mcdonald}}, \bibinfo {author} {\bibfnamefont {F.~F.}\
			\bibnamefont {Balakirev}}, \bibinfo {author} {\bibfnamefont {M.}~\bibnamefont
			{Jaime}}, \emph {et~al.},\ }\href {https://doi.org/10.1038/nature23315}
	{\bibfield  {journal} {\bibinfo  {journal} {Nature}\ }\textbf {\bibinfo
			{volume} {548}},\ \bibinfo {pages} {313} (\bibinfo {year}
		{2017})}\BibitemShut {NoStop}%
	\bibitem [{\citenamefont {Hegger}\ \emph {et~al.}(2000)\citenamefont {Hegger},
		\citenamefont {Petrovic}, \citenamefont {Moshopoulou}, \citenamefont
		{Hundley}, \citenamefont {Sarrao}, \citenamefont {Fisk},\ and\ \citenamefont
		{Thompson}}]{Hegger_2000}%
	\BibitemOpen
	\bibfield  {author} {\bibinfo {author} {\bibfnamefont {H.}~\bibnamefont
			{Hegger}}, \bibinfo {author} {\bibfnamefont {C.}~\bibnamefont {Petrovic}},
		\bibinfo {author} {\bibfnamefont {E.~G.}\ \bibnamefont {Moshopoulou}},
		\bibinfo {author} {\bibfnamefont {M.~F.}\ \bibnamefont {Hundley}}, \bibinfo
		{author} {\bibfnamefont {J.~L.}\ \bibnamefont {Sarrao}}, \bibinfo {author}
		{\bibfnamefont {Z.}~\bibnamefont {Fisk}},\ and\ \bibinfo {author}
		{\bibfnamefont {J.~D.}\ \bibnamefont {Thompson}},\ }\href
	{https://doi.org/10.1103/PhysRevLett.84.4986} {\bibfield  {journal} {\bibinfo
			{journal} {Phys. Rev. Lett.}\ }\textbf {\bibinfo {volume} {84}},\ \bibinfo
		{pages} {4986} (\bibinfo {year} {2000})}\BibitemShut {NoStop}%
	\bibitem [{\citenamefont {Cornelius}\ \emph {et~al.}(2001)\citenamefont
		{Cornelius}, \citenamefont {Pagliuso}, \citenamefont {Hundley},\ and\
		\citenamefont {Sarrao}}]{Cornelius_2001}%
	\BibitemOpen
	\bibfield  {author} {\bibinfo {author} {\bibfnamefont {A.~L.}\ \bibnamefont
			{Cornelius}}, \bibinfo {author} {\bibfnamefont {P.~G.}\ \bibnamefont
			{Pagliuso}}, \bibinfo {author} {\bibfnamefont {M.~F.}\ \bibnamefont
			{Hundley}},\ and\ \bibinfo {author} {\bibfnamefont {J.~L.}\ \bibnamefont
			{Sarrao}},\ }\href {https://doi.org/10.1103/PhysRevB.64.144411} {\bibfield
		{journal} {\bibinfo  {journal} {Phys. Rev. B}\ }\textbf {\bibinfo {volume}
			{64}},\ \bibinfo {pages} {144411} (\bibinfo {year} {2001})}\BibitemShut
	{NoStop}%
	\bibitem [{\citenamefont {Ueda}\ \emph {et~al.}(2004)\citenamefont {Ueda},
		\citenamefont {Shishido}, \citenamefont {Hashimoto}, \citenamefont {Okubo},
		\citenamefont {Yamada}, \citenamefont {Inada}, \citenamefont {Settai},
		\citenamefont {Harima}, \citenamefont {Galatanu}, \citenamefont {Yamamoto},
		\citenamefont {Nakamura}, \citenamefont {Sugiyama}, \citenamefont {Takeuchi},
		\citenamefont {Kindo}, \citenamefont {Namiki}, \citenamefont {Aoki},
		\citenamefont {Sato},\ and\ \citenamefont {\=Onuki}}]{Ueda_2004}%
	\BibitemOpen
	\bibfield  {author} {\bibinfo {author} {\bibfnamefont {T.}~\bibnamefont
			{Ueda}}, \bibinfo {author} {\bibfnamefont {H.}~\bibnamefont {Shishido}},
		\bibinfo {author} {\bibfnamefont {S.}~\bibnamefont {Hashimoto}}, \bibinfo
		{author} {\bibfnamefont {T.}~\bibnamefont {Okubo}}, \bibinfo {author}
		{\bibfnamefont {M.}~\bibnamefont {Yamada}}, \bibinfo {author} {\bibfnamefont
			{Y.}~\bibnamefont {Inada}}, \bibinfo {author} {\bibfnamefont
			{R.}~\bibnamefont {Settai}}, \bibinfo {author} {\bibfnamefont
			{H.}~\bibnamefont {Harima}}, \bibinfo {author} {\bibfnamefont
			{A.}~\bibnamefont {Galatanu}}, \bibinfo {author} {\bibfnamefont
			{E.}~\bibnamefont {Yamamoto}}, \bibinfo {author} {\bibfnamefont
			{N.}~\bibnamefont {Nakamura}}, \bibinfo {author} {\bibfnamefont
			{K.}~\bibnamefont {Sugiyama}}, \bibinfo {author} {\bibfnamefont
			{T.}~\bibnamefont {Takeuchi}}, \bibinfo {author} {\bibfnamefont
			{K.}~\bibnamefont {Kindo}}, \bibinfo {author} {\bibfnamefont
			{T.}~\bibnamefont {Namiki}}, \bibinfo {author} {\bibfnamefont
			{Y.}~\bibnamefont {Aoki}}, \bibinfo {author} {\bibfnamefont {H.}~\bibnamefont
			{Sato}},\ and\ \bibinfo {author} {\bibfnamefont {Y.}~\bibnamefont
			{\=Onuki}},\ }\href {https://doi.org/10.1143/JPSJ.73.649} {\bibfield
		{journal} {\bibinfo  {journal} {Journal of the Physical Society of Japan}\
		}\textbf {\bibinfo {volume} {73}},\ \bibinfo {pages} {649} (\bibinfo {year}
		{2004})}\BibitemShut {NoStop}%
	\bibitem [{\citenamefont {Yashima}\ \emph {et~al.}(2010)\citenamefont
		{Yashima}, \citenamefont {Taniguchi}, \citenamefont {Miyazaki}, \citenamefont
		{Mukuda}, \citenamefont {Kitaoka}, \citenamefont {Shishido}, \citenamefont
		{Settai},\ and\ \citenamefont {\=Onuki}}]{Yashima_2010}%
	\BibitemOpen
	\bibfield  {author} {\bibinfo {author} {\bibfnamefont {M.}~\bibnamefont
			{Yashima}}, \bibinfo {author} {\bibfnamefont {S.}~\bibnamefont {Taniguchi}},
		\bibinfo {author} {\bibfnamefont {H.}~\bibnamefont {Miyazaki}}, \bibinfo
		{author} {\bibfnamefont {H.}~\bibnamefont {Mukuda}}, \bibinfo {author}
		{\bibfnamefont {Y.}~\bibnamefont {Kitaoka}}, \bibinfo {author} {\bibfnamefont
			{H.}~\bibnamefont {Shishido}}, \bibinfo {author} {\bibfnamefont
			{R.}~\bibnamefont {Settai}},\ and\ \bibinfo {author} {\bibfnamefont
			{Y.}~\bibnamefont {\=Onuki}},\ }\href
	{https://doi.org/10.1088/1742-6596/200/1/012238} {\bibfield  {journal}
		{\bibinfo  {journal} {Journal of Physics: Conference Series}\ }\textbf
		{\bibinfo {volume} {200}},\ \bibinfo {pages} {012238} (\bibinfo {year}
		{2010})}\BibitemShut {NoStop}%
	\bibitem [{\citenamefont {Bauer}\ \emph {et~al.}(2010)\citenamefont {Bauer},
		\citenamefont {Lee}, \citenamefont {Sidorov}, \citenamefont {Kurita},
		\citenamefont {Gofryk}, \citenamefont {Zhu}, \citenamefont {Ronning},
		\citenamefont {Movshovich}, \citenamefont {Thompson},\ and\ \citenamefont
		{Park}}]{Bauer_2010_pressure}%
	\BibitemOpen
	\bibfield  {author} {\bibinfo {author} {\bibfnamefont {E.~D.}\ \bibnamefont
			{Bauer}}, \bibinfo {author} {\bibfnamefont {H.~O.}\ \bibnamefont {Lee}},
		\bibinfo {author} {\bibfnamefont {V.~A.}\ \bibnamefont {Sidorov}}, \bibinfo
		{author} {\bibfnamefont {N.}~\bibnamefont {Kurita}}, \bibinfo {author}
		{\bibfnamefont {K.}~\bibnamefont {Gofryk}}, \bibinfo {author} {\bibfnamefont
			{J.-X.}\ \bibnamefont {Zhu}}, \bibinfo {author} {\bibfnamefont
			{F.}~\bibnamefont {Ronning}}, \bibinfo {author} {\bibfnamefont
			{R.}~\bibnamefont {Movshovich}}, \bibinfo {author} {\bibfnamefont {J.~D.}\
			\bibnamefont {Thompson}},\ and\ \bibinfo {author} {\bibfnamefont
			{T.}~\bibnamefont {Park}},\ }\href
	{https://doi.org/10.1103/PhysRevB.81.180507} {\bibfield  {journal} {\bibinfo
			{journal} {Phys. Rev. B}\ }\textbf {\bibinfo {volume} {81}},\ \bibinfo
		{pages} {180507} (\bibinfo {year} {2010})}\BibitemShut {NoStop}%
	\bibitem [{\citenamefont {Honda}\ \emph {et~al.}(2008)\citenamefont {Honda},
		\citenamefont {Measson}, \citenamefont {Nakano}, \citenamefont {Yoshitani},
		\citenamefont {Yamamoto}, \citenamefont {Haga}, \citenamefont {Takeuchi},
		\citenamefont {Yamagami}, \citenamefont {Shimizu}, \citenamefont {Settai},\
		and\ \citenamefont {\=Onuki}}]{Honda_2008_Pressure}%
	\BibitemOpen
	\bibfield  {author} {\bibinfo {author} {\bibfnamefont {F.}~\bibnamefont
			{Honda}}, \bibinfo {author} {\bibfnamefont {M.-A.}\ \bibnamefont {Measson}},
		\bibinfo {author} {\bibfnamefont {Y.}~\bibnamefont {Nakano}}, \bibinfo
		{author} {\bibfnamefont {N.}~\bibnamefont {Yoshitani}}, \bibinfo {author}
		{\bibfnamefont {E.}~\bibnamefont {Yamamoto}}, \bibinfo {author}
		{\bibfnamefont {Y.}~\bibnamefont {Haga}}, \bibinfo {author} {\bibfnamefont
			{T.}~\bibnamefont {Takeuchi}}, \bibinfo {author} {\bibfnamefont
			{H.}~\bibnamefont {Yamagami}}, \bibinfo {author} {\bibfnamefont
			{K.}~\bibnamefont {Shimizu}}, \bibinfo {author} {\bibfnamefont
			{R.}~\bibnamefont {Settai}},\ and\ \bibinfo {author} {\bibfnamefont
			{Y.}~\bibnamefont {\=Onuki}},\ }\href
	{https://doi.org/10.1143/JPSJ.77.043701} {\bibfield  {journal} {\bibinfo
			{journal} {Journal of the Physical Society of Japan}\ }\textbf {\bibinfo
			{volume} {77}},\ \bibinfo {pages} {043701} (\bibinfo {year}
		{2008})}\BibitemShut {NoStop}%
	\bibitem [{\citenamefont {Onimaru}\ \emph {et~al.}(2008)\citenamefont
		{Onimaru}, \citenamefont {F.~Inoue}, \citenamefont {Shigetoh}, \citenamefont
		{Umeo}, \citenamefont {Kubo}, \citenamefont {A.~Ribeiro}, \citenamefont
		{Ishida}, \citenamefont {A.~Avila}, \citenamefont {Ohoyama}, \citenamefont
		{Sera},\ and\ \citenamefont {Takabatake}}]{Onimaru_2008}%
	\BibitemOpen
	\bibfield  {author} {\bibinfo {author} {\bibfnamefont {T.}~\bibnamefont
			{Onimaru}}, \bibinfo {author} {\bibfnamefont {Y.}~\bibnamefont {F.~Inoue}},
		\bibinfo {author} {\bibfnamefont {K.}~\bibnamefont {Shigetoh}}, \bibinfo
		{author} {\bibfnamefont {K.}~\bibnamefont {Umeo}}, \bibinfo {author}
		{\bibfnamefont {H.}~\bibnamefont {Kubo}}, \bibinfo {author} {\bibfnamefont
			{R.}~\bibnamefont {A.~Ribeiro}}, \bibinfo {author} {\bibfnamefont
			{A.}~\bibnamefont {Ishida}}, \bibinfo {author} {\bibfnamefont
			{M.}~\bibnamefont {A.~Avila}}, \bibinfo {author} {\bibfnamefont
			{K.}~\bibnamefont {Ohoyama}}, \bibinfo {author} {\bibfnamefont
			{M.}~\bibnamefont {Sera}},\ and\ \bibinfo {author} {\bibfnamefont
			{T.}~\bibnamefont {Takabatake}},\ }\href
	{https://doi.org/10.1143/JPSJ.77.074708} {\bibfield  {journal} {\bibinfo
			{journal} {Journal of the Physical Society of Japan}\ }\textbf {\bibinfo
			{volume} {77}},\ \bibinfo {pages} {074708} (\bibinfo {year}
		{2008})}\BibitemShut {NoStop}%
	\bibitem [{\citenamefont {Steglich}\ \emph {et~al.}(1996)\citenamefont
		{Steglich}, \citenamefont {Buschinger}, \citenamefont {Gegenwart},
		\citenamefont {Lohmann}, \citenamefont {Helfrich}, \citenamefont
		{Langhammer}, \citenamefont {Hellmann}, \citenamefont {Donnevert},
		\citenamefont {Thomas}, \citenamefont {Link}, \citenamefont {Geibel},
		\citenamefont {Lang}, \citenamefont {Sparn},\ and\ \citenamefont
		{Assmus}}]{Steglich_1996}%
	\BibitemOpen
	\bibfield  {author} {\bibinfo {author} {\bibfnamefont {F.}~\bibnamefont
			{Steglich}}, \bibinfo {author} {\bibfnamefont {B.}~\bibnamefont
			{Buschinger}}, \bibinfo {author} {\bibfnamefont {P.}~\bibnamefont
			{Gegenwart}}, \bibinfo {author} {\bibfnamefont {M.}~\bibnamefont {Lohmann}},
		\bibinfo {author} {\bibfnamefont {R.}~\bibnamefont {Helfrich}}, \bibinfo
		{author} {\bibfnamefont {C.}~\bibnamefont {Langhammer}}, \bibinfo {author}
		{\bibfnamefont {P.}~\bibnamefont {Hellmann}}, \bibinfo {author}
		{\bibfnamefont {L.}~\bibnamefont {Donnevert}}, \bibinfo {author}
		{\bibfnamefont {S.}~\bibnamefont {Thomas}}, \bibinfo {author} {\bibfnamefont
			{A.}~\bibnamefont {Link}}, \bibinfo {author} {\bibfnamefont {C.}~\bibnamefont
			{Geibel}}, \bibinfo {author} {\bibfnamefont {M.}~\bibnamefont {Lang}},
		\bibinfo {author} {\bibfnamefont {G.}~\bibnamefont {Sparn}},\ and\ \bibinfo
		{author} {\bibfnamefont {W.}~\bibnamefont {Assmus}},\ }\href
	{https://doi.org/10.1088/0953-8984/8/48/016} {\bibfield  {journal} {\bibinfo
			{journal} {Journal of Physics: Condensed Matter}\ }\textbf {\bibinfo {volume}
			{8}},\ \bibinfo {pages} {9909} (\bibinfo {year} {1996})}\BibitemShut
	{NoStop}%
	\bibitem [{\citenamefont {Sheikin}\ \emph {et~al.}(2002)\citenamefont
		{Sheikin}, \citenamefont {Wang}, \citenamefont {Bouquet}, \citenamefont
		{Lejay},\ and\ \citenamefont {Junod}}]{Sheikin_2002}%
	\BibitemOpen
	\bibfield  {author} {\bibinfo {author} {\bibfnamefont {I.}~\bibnamefont
			{Sheikin}}, \bibinfo {author} {\bibfnamefont {Y.}~\bibnamefont {Wang}},
		\bibinfo {author} {\bibfnamefont {F.}~\bibnamefont {Bouquet}}, \bibinfo
		{author} {\bibfnamefont {P.}~\bibnamefont {Lejay}},\ and\ \bibinfo {author}
		{\bibfnamefont {A.}~\bibnamefont {Junod}},\ }\href
	{https://doi.org/10.1088/0953-8984/14/28/104} {\bibfield  {journal} {\bibinfo
			{journal} {Journal of Physics: Condensed Matter}\ }\textbf {\bibinfo
			{volume} {14}},\ \bibinfo {pages} {L543} (\bibinfo {year}
		{2002})}\BibitemShut {NoStop}%
	\bibitem [{\citenamefont {Demuer}\ \emph {et~al.}(2001)\citenamefont {Demuer},
		\citenamefont {Jaccard}, \citenamefont {Sheikin}, \citenamefont {Raymond},
		\citenamefont {Salce}, \citenamefont {Thomasson}, \citenamefont
		{Braithwaite},\ and\ \citenamefont {Flouquet}}]{Demuer_2001}%
	\BibitemOpen
	\bibfield  {author} {\bibinfo {author} {\bibfnamefont {A.}~\bibnamefont
			{Demuer}}, \bibinfo {author} {\bibfnamefont {D.}~\bibnamefont {Jaccard}},
		\bibinfo {author} {\bibfnamefont {I.}~\bibnamefont {Sheikin}}, \bibinfo
		{author} {\bibfnamefont {S.}~\bibnamefont {Raymond}}, \bibinfo {author}
		{\bibfnamefont {B.}~\bibnamefont {Salce}}, \bibinfo {author} {\bibfnamefont
			{J.}~\bibnamefont {Thomasson}}, \bibinfo {author} {\bibfnamefont
			{D.}~\bibnamefont {Braithwaite}},\ and\ \bibinfo {author} {\bibfnamefont
			{J.}~\bibnamefont {Flouquet}},\ }\href
	{https://doi.org/10.1088/0953-8984/13/41/320} {\bibfield  {journal} {\bibinfo
			{journal} {Journal of Physics: Condensed Matter}\ }\textbf {\bibinfo
			{volume} {13}},\ \bibinfo {pages} {9335} (\bibinfo {year}
		{2001})}\BibitemShut {NoStop}%
	\bibitem [{\citenamefont {Movshovich}\ \emph {et~al.}(1996)\citenamefont
		{Movshovich}, \citenamefont {Graf}, \citenamefont {Mandrus}, \citenamefont
		{Hundley}, \citenamefont {Thompson}, \citenamefont {Fisher}, \citenamefont
		{Phillips},\ and\ \citenamefont {Smith}}]{MOVSHOVICH1996126}%
	\BibitemOpen
	\bibfield  {author} {\bibinfo {author} {\bibfnamefont {R.}~\bibnamefont
			{Movshovich}}, \bibinfo {author} {\bibfnamefont {T.}~\bibnamefont {Graf}},
		\bibinfo {author} {\bibfnamefont {D.}~\bibnamefont {Mandrus}}, \bibinfo
		{author} {\bibfnamefont {M.}~\bibnamefont {Hundley}}, \bibinfo {author}
		{\bibfnamefont {J.}~\bibnamefont {Thompson}}, \bibinfo {author}
		{\bibfnamefont {R.}~\bibnamefont {Fisher}}, \bibinfo {author} {\bibfnamefont
			{N.}~\bibnamefont {Phillips}},\ and\ \bibinfo {author} {\bibfnamefont
			{J.}~\bibnamefont {Smith}},\ }\href
	{https://doi.org/https://doi.org/10.1016/0921-4526(96)00058-0} {\bibfield
		{journal} {\bibinfo  {journal} {Physica B: Condensed Matter}\ }\textbf
		{\bibinfo {volume} {223-224}},\ \bibinfo {pages} {126} (\bibinfo {year}
		{1996})},\ \bibinfo {note} {proceedings of the International Conference on
		Strongly Correlated Electron Systems}\BibitemShut {NoStop}%
	\bibitem [{\citenamefont {Quezel}\ \emph {et~al.}(1984)\citenamefont {Quezel},
		\citenamefont {Rossat-Mignod}, \citenamefont {Chevalier}, \citenamefont
		{Lejay},\ and\ \citenamefont {Etourneau}}]{QUEZEL1984685}%
	\BibitemOpen
	\bibfield  {author} {\bibinfo {author} {\bibfnamefont {S.}~\bibnamefont
			{Quezel}}, \bibinfo {author} {\bibfnamefont {J.}~\bibnamefont
			{Rossat-Mignod}}, \bibinfo {author} {\bibfnamefont {B.}~\bibnamefont
			{Chevalier}}, \bibinfo {author} {\bibfnamefont {P.}~\bibnamefont {Lejay}},\
		and\ \bibinfo {author} {\bibfnamefont {J.}~\bibnamefont {Etourneau}},\ }\href
	{https://doi.org/https://doi.org/10.1016/0038-1098(84)90221-7} {\bibfield
		{journal} {\bibinfo  {journal} {Solid State Communications}\ }\textbf
		{\bibinfo {volume} {49}},\ \bibinfo {pages} {685} (\bibinfo {year}
		{1984})}\BibitemShut {NoStop}%
	\bibitem [{\citenamefont {Graf}\ \emph {et~al.}(1998)\citenamefont {Graf},
		\citenamefont {Hundley}, \citenamefont {Modler}, \citenamefont {Movshovich},
		\citenamefont {Thompson}, \citenamefont {Mandrus}, \citenamefont {Fisher},\
		and\ \citenamefont {Phillips}}]{Graf_1998}%
	\BibitemOpen
	\bibfield  {author} {\bibinfo {author} {\bibfnamefont {T.}~\bibnamefont
			{Graf}}, \bibinfo {author} {\bibfnamefont {M.~F.}\ \bibnamefont {Hundley}},
		\bibinfo {author} {\bibfnamefont {R.}~\bibnamefont {Modler}}, \bibinfo
		{author} {\bibfnamefont {R.}~\bibnamefont {Movshovich}}, \bibinfo {author}
		{\bibfnamefont {J.~D.}\ \bibnamefont {Thompson}}, \bibinfo {author}
		{\bibfnamefont {D.}~\bibnamefont {Mandrus}}, \bibinfo {author} {\bibfnamefont
			{R.~A.}\ \bibnamefont {Fisher}},\ and\ \bibinfo {author} {\bibfnamefont
			{N.~E.}\ \bibnamefont {Phillips}},\ }\href
	{https://doi.org/10.1103/PhysRevB.57.7442} {\bibfield  {journal} {\bibinfo
			{journal} {Phys. Rev. B}\ }\textbf {\bibinfo {volume} {57}},\ \bibinfo
		{pages} {7442} (\bibinfo {year} {1998})}\BibitemShut {NoStop}%
	\bibitem [{\citenamefont {Klaasse}\ \emph {et~al.}(1987)\citenamefont
		{Klaasse}, \citenamefont {Veenhuizen}, \citenamefont {B{\"o}HM},
		\citenamefont {Bredl}, \citenamefont {Gottwick}, \citenamefont {Mayer},
		\citenamefont {Pawlak}, \citenamefont {Rauchschwalbe}, \citenamefont
		{Spille}, \citenamefont {Steglich} \emph {et~al.}}]{DEBOER198791}%
	\BibitemOpen
	\bibfield  {author} {\bibinfo {author} {\bibfnamefont {J.}~\bibnamefont
			{Klaasse}}, \bibinfo {author} {\bibfnamefont {P.}~\bibnamefont {Veenhuizen}},
		\bibinfo {author} {\bibfnamefont {A.}~\bibnamefont {B{\"o}HM}}, \bibinfo
		{author} {\bibfnamefont {C.}~\bibnamefont {Bredl}}, \bibinfo {author}
		{\bibfnamefont {U.}~\bibnamefont {Gottwick}}, \bibinfo {author}
		{\bibfnamefont {H.}~\bibnamefont {Mayer}}, \bibinfo {author} {\bibfnamefont
			{L.}~\bibnamefont {Pawlak}}, \bibinfo {author} {\bibfnamefont
			{U.}~\bibnamefont {Rauchschwalbe}}, \bibinfo {author} {\bibfnamefont
			{H.}~\bibnamefont {Spille}}, \bibinfo {author} {\bibfnamefont
			{F.}~\bibnamefont {Steglich}}, \emph {et~al.},\ }in\ \href
	{https://doi.org/https://doi.org/10.1016/B978-1-4832-2948-5.50031-9} {\emph
		{\bibinfo {booktitle} {Anomalous Rare Earths and Actinides}}}\ (\bibinfo
	{publisher} {Elsevier},\ \bibinfo {year} {1987})\ pp.\ \bibinfo {pages}
	{91--94}\BibitemShut {NoStop}%
	\bibitem [{\citenamefont {Fisher}\ \emph {et~al.}(1994)\citenamefont {Fisher},
		\citenamefont {Emerson}, \citenamefont {Caspary}, \citenamefont {Phillips},\
		and\ \citenamefont {Steglich}}]{FISHER1994459}%
	\BibitemOpen
	\bibfield  {author} {\bibinfo {author} {\bibfnamefont {R.}~\bibnamefont
			{Fisher}}, \bibinfo {author} {\bibfnamefont {J.}~\bibnamefont {Emerson}},
		\bibinfo {author} {\bibfnamefont {R.}~\bibnamefont {Caspary}}, \bibinfo
		{author} {\bibfnamefont {N.}~\bibnamefont {Phillips}},\ and\ \bibinfo
		{author} {\bibfnamefont {F.}~\bibnamefont {Steglich}},\ }\href
	{https://doi.org/https://doi.org/10.1016/0921-4526(94)90559-2} {\bibfield
		{journal} {\bibinfo  {journal} {Physica B: Condensed Matter}\ }\textbf
		{\bibinfo {volume} {194-196}},\ \bibinfo {pages} {459} (\bibinfo {year}
		{1994})}\BibitemShut {NoStop}%
	\bibitem [{\citenamefont {Thamizhavel}\ \emph
		{et~al.}(2005{\natexlab{a}})\citenamefont {Thamizhavel}, \citenamefont
		{Nakashima}, \citenamefont {Obiraki}, \citenamefont {Nakashima},
		\citenamefont {D~Matsuda}, \citenamefont {Haga}, \citenamefont {Sugiyama},
		\citenamefont {Takeuchi}, \citenamefont {Settai}, \citenamefont {Hagiwara},
		\citenamefont {Kindo},\ and\ \citenamefont {\=Onuki}}]{Thamizhavel_2005}%
	\BibitemOpen
	\bibfield  {author} {\bibinfo {author} {\bibfnamefont {A.}~\bibnamefont
			{Thamizhavel}}, \bibinfo {author} {\bibfnamefont {H.}~\bibnamefont
			{Nakashima}}, \bibinfo {author} {\bibfnamefont {Y.}~\bibnamefont {Obiraki}},
		\bibinfo {author} {\bibfnamefont {M.}~\bibnamefont {Nakashima}}, \bibinfo
		{author} {\bibfnamefont {T.}~\bibnamefont {D~Matsuda}}, \bibinfo {author}
		{\bibfnamefont {Y.}~\bibnamefont {Haga}}, \bibinfo {author} {\bibfnamefont
			{K.}~\bibnamefont {Sugiyama}}, \bibinfo {author} {\bibfnamefont
			{T.}~\bibnamefont {Takeuchi}}, \bibinfo {author} {\bibfnamefont
			{R.}~\bibnamefont {Settai}}, \bibinfo {author} {\bibfnamefont
			{M.}~\bibnamefont {Hagiwara}}, \bibinfo {author} {\bibfnamefont
			{K.}~\bibnamefont {Kindo}},\ and\ \bibinfo {author} {\bibfnamefont
			{Y.}~\bibnamefont {\=Onuki}},\ }\href {https://doi.org/10.1143/JPSJ.74.2843}
	{\bibfield  {journal} {\bibinfo  {journal} {Journal of the Physical Society
				of Japan}\ }\textbf {\bibinfo {volume} {74}},\ \bibinfo {pages} {2843}
		(\bibinfo {year} {2005}{\natexlab{a}})}\BibitemShut {NoStop}%
	\bibitem [{\citenamefont {Nakashima}\ \emph {et~al.}(2005)\citenamefont
		{Nakashima}, \citenamefont {Kohara}, \citenamefont {Thamizhavel},
		\citenamefont {Matsuda}, \citenamefont {Haga}, \citenamefont {Hedo},
		\citenamefont {Uwatoko}, \citenamefont {Settai},\ and\ \citenamefont
		{\=Onuki}}]{Nakashima_2005}%
	\BibitemOpen
	\bibfield  {author} {\bibinfo {author} {\bibfnamefont {M.}~\bibnamefont
			{Nakashima}}, \bibinfo {author} {\bibfnamefont {H.}~\bibnamefont {Kohara}},
		\bibinfo {author} {\bibfnamefont {A.}~\bibnamefont {Thamizhavel}}, \bibinfo
		{author} {\bibfnamefont {T.~D.}\ \bibnamefont {Matsuda}}, \bibinfo {author}
		{\bibfnamefont {Y.}~\bibnamefont {Haga}}, \bibinfo {author} {\bibfnamefont
			{M.}~\bibnamefont {Hedo}}, \bibinfo {author} {\bibfnamefont {Y.}~\bibnamefont
			{Uwatoko}}, \bibinfo {author} {\bibfnamefont {R.}~\bibnamefont {Settai}},\
		and\ \bibinfo {author} {\bibfnamefont {Y.}~\bibnamefont {\=Onuki}},\ }\href
	{https://doi.org/10.1088/0953-8984/17/28/012} {\bibfield  {journal} {\bibinfo
			{journal} {Journal of Physics: Condensed Matter}\ }\textbf {\bibinfo
			{volume} {17}},\ \bibinfo {pages} {4539} (\bibinfo {year}
		{2005})}\BibitemShut {NoStop}%
	\bibitem [{\citenamefont {Mun}\ \emph {et~al.}(2010)\citenamefont {Mun},
		\citenamefont {Bud'ko}, \citenamefont {Kreyssig},\ and\ \citenamefont
		{Canfield}}]{Mun_2010_Tuning}%
	\BibitemOpen
	\bibfield  {author} {\bibinfo {author} {\bibfnamefont {E.~D.}\ \bibnamefont
			{Mun}}, \bibinfo {author} {\bibfnamefont {S.~L.}\ \bibnamefont {Bud'ko}},
		\bibinfo {author} {\bibfnamefont {A.}~\bibnamefont {Kreyssig}},\ and\
		\bibinfo {author} {\bibfnamefont {P.~C.}\ \bibnamefont {Canfield}},\ }\href
	{https://doi.org/10.1103/PhysRevB.82.054424} {\bibfield  {journal} {\bibinfo
			{journal} {Phys. Rev. B}\ }\textbf {\bibinfo {volume} {82}},\ \bibinfo
		{pages} {054424} (\bibinfo {year} {2010})}\BibitemShut {NoStop}%
	\bibitem [{\citenamefont {Nakashima}\ \emph {et~al.}(2004)\citenamefont
		{Nakashima}, \citenamefont {Tabata}, \citenamefont {Thamizhavel},
		\citenamefont {Kobayashi}, \citenamefont {Hedo}, \citenamefont {Uwatoko},
		\citenamefont {Shimizu}, \citenamefont {Settai},\ and\ \citenamefont
		{\=Onuki}}]{Nakashima_2004}%
	\BibitemOpen
	\bibfield  {author} {\bibinfo {author} {\bibfnamefont {M.}~\bibnamefont
			{Nakashima}}, \bibinfo {author} {\bibfnamefont {K.}~\bibnamefont {Tabata}},
		\bibinfo {author} {\bibfnamefont {A.}~\bibnamefont {Thamizhavel}}, \bibinfo
		{author} {\bibfnamefont {T.~C.}\ \bibnamefont {Kobayashi}}, \bibinfo {author}
		{\bibfnamefont {M.}~\bibnamefont {Hedo}}, \bibinfo {author} {\bibfnamefont
			{Y.}~\bibnamefont {Uwatoko}}, \bibinfo {author} {\bibfnamefont
			{K.}~\bibnamefont {Shimizu}}, \bibinfo {author} {\bibfnamefont
			{R.}~\bibnamefont {Settai}},\ and\ \bibinfo {author} {\bibfnamefont
			{Y.}~\bibnamefont {\=Onuki}},\ }\href
	{https://doi.org/10.1088/0953-8984/16/20/L01} {\bibfield  {journal} {\bibinfo
			{journal} {Journal of Physics: Condensed Matter}\ }\textbf {\bibinfo
			{volume} {16}},\ \bibinfo {pages} {L255} (\bibinfo {year}
		{2004})}\BibitemShut {NoStop}%
	\bibitem [{\citenamefont {Takeuchi}\ \emph {et~al.}(2007)\citenamefont
		{Takeuchi}, \citenamefont {Yasuda}, \citenamefont {Tsujino}, \citenamefont
		{Shishido}, \citenamefont {Settai}, \citenamefont {Harima},\ and\
		\citenamefont {\=Onuki}}]{Takeuchi_2007}%
	\BibitemOpen
	\bibfield  {author} {\bibinfo {author} {\bibfnamefont {T.}~\bibnamefont
			{Takeuchi}}, \bibinfo {author} {\bibfnamefont {T.}~\bibnamefont {Yasuda}},
		\bibinfo {author} {\bibfnamefont {M.}~\bibnamefont {Tsujino}}, \bibinfo
		{author} {\bibfnamefont {H.}~\bibnamefont {Shishido}}, \bibinfo {author}
		{\bibfnamefont {R.}~\bibnamefont {Settai}}, \bibinfo {author} {\bibfnamefont
			{H.}~\bibnamefont {Harima}},\ and\ \bibinfo {author} {\bibfnamefont
			{Y.}~\bibnamefont {\=Onuki}},\ }\href
	{https://doi.org/10.1143/JPSJ.76.014702} {\bibfield  {journal} {\bibinfo
			{journal} {Journal of the Physical Society of Japan}\ }\textbf {\bibinfo
			{volume} {76}},\ \bibinfo {pages} {014702} (\bibinfo {year}
		{2007})}\BibitemShut {NoStop}%
	\bibitem [{\citenamefont {Kimura}\ \emph {et~al.}(2007)\citenamefont {Kimura},
		\citenamefont {Muro},\ and\ \citenamefont {Aoki}}]{Kimura_2007}%
	\BibitemOpen
	\bibfield  {author} {\bibinfo {author} {\bibfnamefont {N.}~\bibnamefont
			{Kimura}}, \bibinfo {author} {\bibfnamefont {Y.}~\bibnamefont {Muro}},\ and\
		\bibinfo {author} {\bibfnamefont {H.}~\bibnamefont {Aoki}},\ }\href
	{https://doi.org/10.1143/JPSJ.76.051010} {\bibfield  {journal} {\bibinfo
			{journal} {Journal of the Physical Society of Japan}\ }\textbf {\bibinfo
			{volume} {76}},\ \bibinfo {pages} {051010} (\bibinfo {year}
		{2007})}\BibitemShut {NoStop}%
	\bibitem [{\citenamefont {P\'asztorov\'a}\ \emph {et~al.}(2019)\citenamefont
		{P\'asztorov\'a}, \citenamefont {Howell}, \citenamefont {Songvilay},
		\citenamefont {Sarte}, \citenamefont {Rodriguez-Rivera}, \citenamefont
		{Ar\'evalo-L\'opez}, \citenamefont {Schmalzl}, \citenamefont {Schneidewind},
		\citenamefont {Dunsiger}, \citenamefont {Singh}, \citenamefont {Petrovic},
		\citenamefont {Hu},\ and\ \citenamefont {Stock}}]{Patztorova_2019}%
	\BibitemOpen
	\bibfield  {author} {\bibinfo {author} {\bibfnamefont {J.}~\bibnamefont
			{P\'asztorov\'a}}, \bibinfo {author} {\bibfnamefont {A.}~\bibnamefont
			{Howell}}, \bibinfo {author} {\bibfnamefont {M.}~\bibnamefont {Songvilay}},
		\bibinfo {author} {\bibfnamefont {P.~M.}\ \bibnamefont {Sarte}}, \bibinfo
		{author} {\bibfnamefont {J.~A.}\ \bibnamefont {Rodriguez-Rivera}}, \bibinfo
		{author} {\bibfnamefont {A.~M.}\ \bibnamefont {Ar\'evalo-L\'opez}}, \bibinfo
		{author} {\bibfnamefont {K.}~\bibnamefont {Schmalzl}}, \bibinfo {author}
		{\bibfnamefont {A.}~\bibnamefont {Schneidewind}}, \bibinfo {author}
		{\bibfnamefont {S.~R.}\ \bibnamefont {Dunsiger}}, \bibinfo {author}
		{\bibfnamefont {D.~K.}\ \bibnamefont {Singh}}, \bibinfo {author}
		{\bibfnamefont {C.}~\bibnamefont {Petrovic}}, \bibinfo {author}
		{\bibfnamefont {R.}~\bibnamefont {Hu}},\ and\ \bibinfo {author}
		{\bibfnamefont {C.}~\bibnamefont {Stock}},\ }\href
	{https://doi.org/10.1103/PhysRevB.99.125144} {\bibfield  {journal} {\bibinfo
			{journal} {Phys. Rev. B}\ }\textbf {\bibinfo {volume} {99}},\ \bibinfo
		{pages} {125144} (\bibinfo {year} {2019})}\BibitemShut {NoStop}%
	\bibitem [{\citenamefont {Okuda}\ \emph {et~al.}(2007)\citenamefont {Okuda},
		\citenamefont {Miyauchi}, \citenamefont {Ida}, \citenamefont {Takeda},
		\citenamefont {Tonohiro}, \citenamefont {Oduchi}, \citenamefont {Yamada},
		\citenamefont {Duc~Dung}, \citenamefont {D.~Matsuda}, \citenamefont {Haga},
		\citenamefont {Takeuchi}, \citenamefont {Hagiwara}, \citenamefont {Kindo},
		\citenamefont {Harima}, \citenamefont {Sugiyama}, \citenamefont {Settai},\
		and\ \citenamefont {\=Onuki}}]{Okuda_2007}%
	\BibitemOpen
	\bibfield  {author} {\bibinfo {author} {\bibfnamefont {Y.}~\bibnamefont
			{Okuda}}, \bibinfo {author} {\bibfnamefont {Y.}~\bibnamefont {Miyauchi}},
		\bibinfo {author} {\bibfnamefont {Y.}~\bibnamefont {Ida}}, \bibinfo {author}
		{\bibfnamefont {Y.}~\bibnamefont {Takeda}}, \bibinfo {author} {\bibfnamefont
			{C.}~\bibnamefont {Tonohiro}}, \bibinfo {author} {\bibfnamefont
			{Y.}~\bibnamefont {Oduchi}}, \bibinfo {author} {\bibfnamefont
			{T.}~\bibnamefont {Yamada}}, \bibinfo {author} {\bibfnamefont
			{N.}~\bibnamefont {Duc~Dung}}, \bibinfo {author} {\bibfnamefont
			{T.}~\bibnamefont {D.~Matsuda}}, \bibinfo {author} {\bibfnamefont
			{Y.}~\bibnamefont {Haga}}, \bibinfo {author} {\bibfnamefont {T.}~\bibnamefont
			{Takeuchi}}, \bibinfo {author} {\bibfnamefont {M.}~\bibnamefont {Hagiwara}},
		\bibinfo {author} {\bibfnamefont {K.}~\bibnamefont {Kindo}}, \bibinfo
		{author} {\bibfnamefont {H.}~\bibnamefont {Harima}}, \bibinfo {author}
		{\bibfnamefont {K.}~\bibnamefont {Sugiyama}}, \bibinfo {author}
		{\bibfnamefont {R.}~\bibnamefont {Settai}},\ and\ \bibinfo {author}
		{\bibfnamefont {Y.}~\bibnamefont {\=Onuki}},\ }\href
	{https://doi.org/10.1143/JPSJ.76.044708} {\bibfield  {journal} {\bibinfo
			{journal} {Journal of the Physical Society of Japan}\ }\textbf {\bibinfo
			{volume} {76}},\ \bibinfo {pages} {044708} (\bibinfo {year}
		{2007})}\BibitemShut {NoStop}%
	\bibitem [{\citenamefont {Settai}\ \emph {et~al.}(2011)\citenamefont {Settai},
		\citenamefont {Katayama}, \citenamefont {Aoki}, \citenamefont {Sheikin},
		\citenamefont {Knebel}, \citenamefont {Flouquet},\ and\ \citenamefont
		{\=Onuki}}]{Settai_2011}%
	\BibitemOpen
	\bibfield  {author} {\bibinfo {author} {\bibfnamefont {R.}~\bibnamefont
			{Settai}}, \bibinfo {author} {\bibfnamefont {K.}~\bibnamefont {Katayama}},
		\bibinfo {author} {\bibfnamefont {D.}~\bibnamefont {Aoki}}, \bibinfo {author}
		{\bibfnamefont {I.}~\bibnamefont {Sheikin}}, \bibinfo {author} {\bibfnamefont
			{G.}~\bibnamefont {Knebel}}, \bibinfo {author} {\bibfnamefont
			{J.}~\bibnamefont {Flouquet}},\ and\ \bibinfo {author} {\bibfnamefont
			{Y.}~\bibnamefont {\=Onuki}},\ }\href
	{https://doi.org/10.1143/JPSJ.80.094703} {\bibfield  {journal} {\bibinfo
			{journal} {Journal of the Physical Society of Japan}\ }\textbf {\bibinfo
			{volume} {80}},\ \bibinfo {pages} {094703} (\bibinfo {year}
		{2011})}\BibitemShut {NoStop}%
	\bibitem [{\citenamefont {Knebel}\ \emph {et~al.}(2009)\citenamefont {Knebel},
		\citenamefont {Aoki}, \citenamefont {Lapertot}, \citenamefont {Salce},
		\citenamefont {Flouquet}, \citenamefont {Kawai}, \citenamefont {Muranaka},
		\citenamefont {Settai},\ and\ \citenamefont {\=Onuki}}]{Knebel_2009}%
	\BibitemOpen
	\bibfield  {author} {\bibinfo {author} {\bibfnamefont {G.}~\bibnamefont
			{Knebel}}, \bibinfo {author} {\bibfnamefont {D.}~\bibnamefont {Aoki}},
		\bibinfo {author} {\bibfnamefont {G.}~\bibnamefont {Lapertot}}, \bibinfo
		{author} {\bibfnamefont {B.}~\bibnamefont {Salce}}, \bibinfo {author}
		{\bibfnamefont {J.}~\bibnamefont {Flouquet}}, \bibinfo {author}
		{\bibfnamefont {T.}~\bibnamefont {Kawai}}, \bibinfo {author} {\bibfnamefont
			{H.}~\bibnamefont {Muranaka}}, \bibinfo {author} {\bibfnamefont
			{R.}~\bibnamefont {Settai}},\ and\ \bibinfo {author} {\bibfnamefont
			{Y.}~\bibnamefont {\=Onuki}},\ }\href
	{https://doi.org/10.1143/JPSJ.78.074714} {\bibfield  {journal} {\bibinfo
			{journal} {Journal of the Physical Society of Japan}\ }\textbf {\bibinfo
			{volume} {78}},\ \bibinfo {pages} {074714} (\bibinfo {year}
		{2009})}\BibitemShut {NoStop}%
	\bibitem [{\citenamefont {Thamizhavel}\ \emph
		{et~al.}(2005{\natexlab{b}})\citenamefont {Thamizhavel}, \citenamefont
		{Takeuchi}, \citenamefont {D~Matsuda}, \citenamefont {Haga}, \citenamefont
		{Sugiyama}, \citenamefont {Settai},\ and\ \citenamefont
		{\=Onuki}}]{Thamizhavel_2005_Unique}%
	\BibitemOpen
	\bibfield  {author} {\bibinfo {author} {\bibfnamefont {A.}~\bibnamefont
			{Thamizhavel}}, \bibinfo {author} {\bibfnamefont {T.}~\bibnamefont
			{Takeuchi}}, \bibinfo {author} {\bibfnamefont {T.}~\bibnamefont {D~Matsuda}},
		\bibinfo {author} {\bibfnamefont {Y.}~\bibnamefont {Haga}}, \bibinfo {author}
		{\bibfnamefont {K.}~\bibnamefont {Sugiyama}}, \bibinfo {author}
		{\bibfnamefont {R.}~\bibnamefont {Settai}},\ and\ \bibinfo {author}
		{\bibfnamefont {Y.}~\bibnamefont {\=Onuki}},\ }\href
	{https://doi.org/10.1143/JPSJ.74.1858} {\bibfield  {journal} {\bibinfo
			{journal} {Journal of the Physical Society of Japan}\ }\textbf {\bibinfo
			{volume} {74}},\ \bibinfo {pages} {1858} (\bibinfo {year}
		{2005}{\natexlab{b}})}\BibitemShut {NoStop}%
	\bibitem [{\citenamefont {Fritsch}\ \emph {et~al.}(2017)\citenamefont
		{Fritsch}, \citenamefont {Lucas}, \citenamefont {Huesges}, \citenamefont
		{Sakai}, \citenamefont {Kittler}, \citenamefont {Taubenheim}, \citenamefont
		{Woitschach}, \citenamefont {Pedersen}, \citenamefont {Grube}, \citenamefont
		{Schmidt}, \citenamefont {Gegenwart}, \citenamefont {Stockert},\ and\
		\citenamefont {v.~L\"ohneysen}}]{Fritsch_2017}%
	\BibitemOpen
	\bibfield  {author} {\bibinfo {author} {\bibfnamefont {V.}~\bibnamefont
			{Fritsch}}, \bibinfo {author} {\bibfnamefont {S.}~\bibnamefont {Lucas}},
		\bibinfo {author} {\bibfnamefont {Z.}~\bibnamefont {Huesges}}, \bibinfo
		{author} {\bibfnamefont {A.}~\bibnamefont {Sakai}}, \bibinfo {author}
		{\bibfnamefont {W.}~\bibnamefont {Kittler}}, \bibinfo {author} {\bibfnamefont
			{C.}~\bibnamefont {Taubenheim}}, \bibinfo {author} {\bibfnamefont
			{S.}~\bibnamefont {Woitschach}}, \bibinfo {author} {\bibfnamefont
			{B.}~\bibnamefont {Pedersen}}, \bibinfo {author} {\bibfnamefont
			{K.}~\bibnamefont {Grube}}, \bibinfo {author} {\bibfnamefont
			{B.}~\bibnamefont {Schmidt}}, \bibinfo {author} {\bibfnamefont
			{P.}~\bibnamefont {Gegenwart}}, \bibinfo {author} {\bibfnamefont
			{O.}~\bibnamefont {Stockert}},\ and\ \bibinfo {author} {\bibfnamefont
			{H.}~\bibnamefont {v.~L\"ohneysen}},\ }\href
	{https://doi.org/10.1088/1742-6596/807/3/032003} {\bibfield  {journal}
		{\bibinfo  {journal} {Journal of Physics: Conference Series}\ }\textbf
		{\bibinfo {volume} {807}},\ \bibinfo {pages} {032003} (\bibinfo {year}
		{2017})}\BibitemShut {NoStop}%
	\bibitem [{\citenamefont {Majumder}\ \emph {et~al.}(2022)\citenamefont
		{Majumder}, \citenamefont {Gupta}, \citenamefont {Luetkens}, \citenamefont
		{Khasanov}, \citenamefont {Stockert}, \citenamefont {Gegenwart},\ and\
		\citenamefont {Fritsch}}]{Majumder_2022}%
	\BibitemOpen
	\bibfield  {author} {\bibinfo {author} {\bibfnamefont {M.}~\bibnamefont
			{Majumder}}, \bibinfo {author} {\bibfnamefont {R.}~\bibnamefont {Gupta}},
		\bibinfo {author} {\bibfnamefont {H.}~\bibnamefont {Luetkens}}, \bibinfo
		{author} {\bibfnamefont {R.}~\bibnamefont {Khasanov}}, \bibinfo {author}
		{\bibfnamefont {O.}~\bibnamefont {Stockert}}, \bibinfo {author}
		{\bibfnamefont {P.}~\bibnamefont {Gegenwart}},\ and\ \bibinfo {author}
		{\bibfnamefont {V.}~\bibnamefont {Fritsch}},\ }\href
	{https://doi.org/10.1103/PhysRevB.105.L180402} {\bibfield  {journal}
		{\bibinfo  {journal} {Phys. Rev. B}\ }\textbf {\bibinfo {volume} {105}},\
		\bibinfo {pages} {L180402} (\bibinfo {year} {2022})}\BibitemShut {NoStop}%
	\bibitem [{\citenamefont {Matsuoka}\ \emph {et~al.}(2015)\citenamefont
		{Matsuoka}, \citenamefont {Hondo}, \citenamefont {Fujii}, \citenamefont
		{Oshima}, \citenamefont {Sugawara}, \citenamefont {Sakurai}, \citenamefont
		{Ohta}, \citenamefont {Kneidinger}, \citenamefont {Salamakha}, \citenamefont
		{Michor},\ and\ \citenamefont {Bauer}}]{Matsuoka_2015}%
	\BibitemOpen
	\bibfield  {author} {\bibinfo {author} {\bibfnamefont {E.}~\bibnamefont
			{Matsuoka}}, \bibinfo {author} {\bibfnamefont {C.}~\bibnamefont {Hondo}},
		\bibinfo {author} {\bibfnamefont {T.}~\bibnamefont {Fujii}}, \bibinfo
		{author} {\bibfnamefont {A.}~\bibnamefont {Oshima}}, \bibinfo {author}
		{\bibfnamefont {H.}~\bibnamefont {Sugawara}}, \bibinfo {author}
		{\bibfnamefont {T.}~\bibnamefont {Sakurai}}, \bibinfo {author} {\bibfnamefont
			{H.}~\bibnamefont {Ohta}}, \bibinfo {author} {\bibfnamefont {F.}~\bibnamefont
			{Kneidinger}}, \bibinfo {author} {\bibfnamefont {L.}~\bibnamefont
			{Salamakha}}, \bibinfo {author} {\bibfnamefont {H.}~\bibnamefont {Michor}},\
		and\ \bibinfo {author} {\bibfnamefont {E.}~\bibnamefont {Bauer}},\ }\href
	{https://doi.org/10.7566/JPSJ.84.073704} {\bibfield  {journal} {\bibinfo
			{journal} {Journal of the Physical Society of Japan}\ }\textbf {\bibinfo
			{volume} {84}},\ \bibinfo {pages} {073704} (\bibinfo {year}
		{2015})}\BibitemShut {NoStop}%
	\bibitem [{\citenamefont {Kotegawa}\ \emph {et~al.}(2019)\citenamefont
		{Kotegawa}, \citenamefont {Matsuoka}, \citenamefont {Uga}, \citenamefont
		{Takemura}, \citenamefont {Manago}, \citenamefont {Chikuchi}, \citenamefont
		{Sugawara}, \citenamefont {Tou},\ and\ \citenamefont
		{Harima}}]{Kotegawa_2019}%
	\BibitemOpen
	\bibfield  {author} {\bibinfo {author} {\bibfnamefont {H.}~\bibnamefont
			{Kotegawa}}, \bibinfo {author} {\bibfnamefont {E.}~\bibnamefont {Matsuoka}},
		\bibinfo {author} {\bibfnamefont {T.}~\bibnamefont {Uga}}, \bibinfo {author}
		{\bibfnamefont {M.}~\bibnamefont {Takemura}}, \bibinfo {author}
		{\bibfnamefont {M.}~\bibnamefont {Manago}}, \bibinfo {author} {\bibfnamefont
			{N.}~\bibnamefont {Chikuchi}}, \bibinfo {author} {\bibfnamefont
			{H.}~\bibnamefont {Sugawara}}, \bibinfo {author} {\bibfnamefont
			{H.}~\bibnamefont {Tou}},\ and\ \bibinfo {author} {\bibfnamefont
			{H.}~\bibnamefont {Harima}},\ }\href {https://doi.org/10.7566/JPSJ.88.093702}
	{\bibfield  {journal} {\bibinfo  {journal} {Journal of the Physical Society
				of Japan}\ }\textbf {\bibinfo {volume} {88}},\ \bibinfo {pages} {093702}
		(\bibinfo {year} {2019})}\BibitemShut {NoStop}%
	\bibitem [{\citenamefont {Dzsaber}\ \emph {et~al.}(2017)\citenamefont
		{Dzsaber}, \citenamefont {Prochaska}, \citenamefont {Sidorenko},
		\citenamefont {Eguchi}, \citenamefont {Svagera}, \citenamefont {Waas},
		\citenamefont {Prokofiev}, \citenamefont {Si},\ and\ \citenamefont
		{Paschen}}]{Dzsaber_2017}%
	\BibitemOpen
	\bibfield  {author} {\bibinfo {author} {\bibfnamefont {S.}~\bibnamefont
			{Dzsaber}}, \bibinfo {author} {\bibfnamefont {L.}~\bibnamefont {Prochaska}},
		\bibinfo {author} {\bibfnamefont {A.}~\bibnamefont {Sidorenko}}, \bibinfo
		{author} {\bibfnamefont {G.}~\bibnamefont {Eguchi}}, \bibinfo {author}
		{\bibfnamefont {R.}~\bibnamefont {Svagera}}, \bibinfo {author} {\bibfnamefont
			{M.}~\bibnamefont {Waas}}, \bibinfo {author} {\bibfnamefont {A.}~\bibnamefont
			{Prokofiev}}, \bibinfo {author} {\bibfnamefont {Q.}~\bibnamefont {Si}},\ and\
		\bibinfo {author} {\bibfnamefont {S.}~\bibnamefont {Paschen}},\ }\href
	{https://doi.org/10.1103/PhysRevLett.118.246601} {\bibfield  {journal}
		{\bibinfo  {journal} {Phys. Rev. Lett.}\ }\textbf {\bibinfo {volume} {118}},\
		\bibinfo {pages} {246601} (\bibinfo {year} {2017})}\BibitemShut {NoStop}%
	\bibitem [{\citenamefont {Lai}\ \emph {et~al.}(2018)\citenamefont {Lai},
		\citenamefont {Grefe}, \citenamefont {Paschen},\ and\ \citenamefont
		{Si}}]{lai2018weyl}%
	\BibitemOpen
	\bibfield  {author} {\bibinfo {author} {\bibfnamefont {H.-H.}\ \bibnamefont
			{Lai}}, \bibinfo {author} {\bibfnamefont {S.~E.}\ \bibnamefont {Grefe}},
		\bibinfo {author} {\bibfnamefont {S.}~\bibnamefont {Paschen}},\ and\ \bibinfo
		{author} {\bibfnamefont {Q.}~\bibnamefont {Si}},\ }\href
	{https://doi.org/10.1073/pnas.1715851115} {\bibfield  {journal} {\bibinfo
			{journal} {Proceedings of the National Academy of Sciences}\ }\textbf
		{\bibinfo {volume} {115}},\ \bibinfo {pages} {93} (\bibinfo {year}
		{2018})}\BibitemShut {NoStop}%
	\bibitem [{\citenamefont {Chen}\ \emph {et~al.}(2022)\citenamefont {Chen},
		\citenamefont {Setty}, \citenamefont {Hu}, \citenamefont {Vergniory},
		\citenamefont {Grefe}, \citenamefont {Fischer}, \citenamefont {Yan},
		\citenamefont {Prokofiev}, \citenamefont {Paschen}, \citenamefont {Cano},\
		and\ \citenamefont {Si}}]{Chen2022}%
	\BibitemOpen
	\bibfield  {author} {\bibinfo {author} {\bibfnamefont {L.}~\bibnamefont
			{Chen}}, \bibinfo {author} {\bibfnamefont {C.}~\bibnamefont {Setty}},
		\bibinfo {author} {\bibfnamefont {H.}~\bibnamefont {Hu}}, \bibinfo {author}
		{\bibfnamefont {M.~G.}\ \bibnamefont {Vergniory}}, \bibinfo {author}
		{\bibfnamefont {S.~E.}\ \bibnamefont {Grefe}}, \bibinfo {author}
		{\bibfnamefont {L.}~\bibnamefont {Fischer}}, \bibinfo {author} {\bibfnamefont
			{G.}~\bibnamefont {Yan}, \bibfnamefont {Xinlinand~Eguchi}}, \bibinfo {author}
		{\bibfnamefont {A.}~\bibnamefont {Prokofiev}}, \bibinfo {author}
		{\bibfnamefont {S.}~\bibnamefont {Paschen}}, \bibinfo {author} {\bibfnamefont
			{J.}~\bibnamefont {Cano}},\ and\ \bibinfo {author} {\bibfnamefont
			{Q.}~\bibnamefont {Si}},\ }\href {https://doi.org/10.1038/s41567-022-01743-4}
	{\bibfield  {journal} {\bibinfo  {journal} {Nature Physics}\ }\textbf
		{\bibinfo {volume} {18}},\ \bibinfo {pages} {1341} (\bibinfo {year}
		{2022})}\BibitemShut {NoStop}%
	\bibitem [{\citenamefont {Chubukov}\ \emph {et~al.}(2006)\citenamefont
		{Chubukov}, \citenamefont {Maslov},\ and\ \citenamefont
		{Millis}}]{ChubukovPRB2006}%
	\BibitemOpen
	\bibfield  {author} {\bibinfo {author} {\bibfnamefont {A.~V.}\ \bibnamefont
			{Chubukov}}, \bibinfo {author} {\bibfnamefont {D.~L.}\ \bibnamefont
			{Maslov}},\ and\ \bibinfo {author} {\bibfnamefont {A.~J.}\ \bibnamefont
			{Millis}},\ }\href {https://doi.org/10.1103/PhysRevB.73.045128} {\bibfield
		{journal} {\bibinfo  {journal} {Phys. Rev. B}\ }\textbf {\bibinfo {volume}
			{73}},\ \bibinfo {pages} {045128} (\bibinfo {year} {2006})}\BibitemShut
	{NoStop}%
	\bibitem [{\citenamefont {Mishra}\ \emph {et~al.}(2021)\citenamefont {Mishra},
		\citenamefont {Demuer}, \citenamefont {Aoki},\ and\ \citenamefont
		{Sheikin}}]{Mishra_2021}%
	\BibitemOpen
	\bibfield  {author} {\bibinfo {author} {\bibfnamefont {S.}~\bibnamefont
			{Mishra}}, \bibinfo {author} {\bibfnamefont {A.}~\bibnamefont {Demuer}},
		\bibinfo {author} {\bibfnamefont {D.}~\bibnamefont {Aoki}},\ and\ \bibinfo
		{author} {\bibfnamefont {I.}~\bibnamefont {Sheikin}},\ }\href
	{https://doi.org/10.1103/PhysRevB.103.045110} {\bibfield  {journal} {\bibinfo
			{journal} {Phys. Rev. B}\ }\textbf {\bibinfo {volume} {103}},\ \bibinfo
		{pages} {045110} (\bibinfo {year} {2021})}\BibitemShut {NoStop}%
	\bibitem [{\citenamefont {Platzman}\ and\ \citenamefont
		{Wolff}(1967)}]{Platzman_1967}%
	\BibitemOpen
	\bibfield  {author} {\bibinfo {author} {\bibfnamefont {P.~M.}\ \bibnamefont
			{Platzman}}\ and\ \bibinfo {author} {\bibfnamefont {P.~A.}\ \bibnamefont
			{Wolff}},\ }\href {https://doi.org/10.1103/PhysRevLett.18.280} {\bibfield
		{journal} {\bibinfo  {journal} {Phys. Rev. Lett.}\ }\textbf {\bibinfo
			{volume} {18}},\ \bibinfo {pages} {280} (\bibinfo {year} {1967})}\BibitemShut
	{NoStop}%
	\bibitem [{\citenamefont {Kushwaha}\ \emph {et~al.}(2019)\citenamefont
		{Kushwaha}, \citenamefont {Chan}, \citenamefont {Park}, \citenamefont
		{Thomas}, \citenamefont {Bauer}, \citenamefont {Thompson}, \citenamefont
		{Ronning}, \citenamefont {Rosa},\ and\ \citenamefont
		{Harrison}}]{Kushwaha2019}%
	\BibitemOpen
	\bibfield  {author} {\bibinfo {author} {\bibfnamefont {S.~K.}\ \bibnamefont
			{Kushwaha}}, \bibinfo {author} {\bibfnamefont {M.~K.}\ \bibnamefont {Chan}},
		\bibinfo {author} {\bibfnamefont {J.}~\bibnamefont {Park}}, \bibinfo {author}
		{\bibfnamefont {S.~M.}\ \bibnamefont {Thomas}}, \bibinfo {author}
		{\bibfnamefont {E.~D.}\ \bibnamefont {Bauer}}, \bibinfo {author}
		{\bibfnamefont {J.~D.}\ \bibnamefont {Thompson}}, \bibinfo {author}
		{\bibfnamefont {F.}~\bibnamefont {Ronning}}, \bibinfo {author} {\bibfnamefont
			{P.~F.~S.}\ \bibnamefont {Rosa}},\ and\ \bibinfo {author} {\bibfnamefont
			{N.}~\bibnamefont {Harrison}},\ }\href
	{https://doi.org/10.1038/s41467-019-13421-w} {\bibfield  {journal} {\bibinfo
			{journal} {Nature Communications}\ }\textbf {\bibinfo {volume} {10}},\
		\bibinfo {pages} {5487} (\bibinfo {year} {2019})}\BibitemShut {NoStop}%
	\bibitem [{\citenamefont {Wartenbe}\ \emph {et~al.}(2022)\citenamefont
		{Wartenbe}, \citenamefont {Tobash}, \citenamefont {Singleton}, \citenamefont
		{Winter}, \citenamefont {Richmond},\ and\ \citenamefont
		{Harrison}}]{Wartenbe_2022}%
	\BibitemOpen
	\bibfield  {author} {\bibinfo {author} {\bibfnamefont {M.}~\bibnamefont
			{Wartenbe}}, \bibinfo {author} {\bibfnamefont {P.~H.}\ \bibnamefont
			{Tobash}}, \bibinfo {author} {\bibfnamefont {J.}~\bibnamefont {Singleton}},
		\bibinfo {author} {\bibfnamefont {L.~E.}\ \bibnamefont {Winter}}, \bibinfo
		{author} {\bibfnamefont {S.}~\bibnamefont {Richmond}},\ and\ \bibinfo
		{author} {\bibfnamefont {N.}~\bibnamefont {Harrison}},\ }\href
	{https://doi.org/10.1103/PhysRevB.105.L041107} {\bibfield  {journal}
		{\bibinfo  {journal} {Phys. Rev. B}\ }\textbf {\bibinfo {volume} {105}},\
		\bibinfo {pages} {L041107} (\bibinfo {year} {2022})}\BibitemShut {NoStop}%
\end{thebibliography}

\begin{thebibliography}{10}%
\makeatletter
\providecommand \@ifxundefined [1]{%
 \@ifx{#1\undefined}
}%
\providecommand \@ifnum [1]{%
 \ifnum #1\expandafter \@firstoftwo
 \else \expandafter \@secondoftwo
 \fi
}%
\providecommand \@ifx [1]{%
 \ifx #1\expandafter \@firstoftwo
 \else \expandafter \@secondoftwo
 \fi
}%
\providecommand \natexlab [1]{#1}%
\providecommand \enquote  [1]{``#1''}%
\providecommand \bibnamefont  [1]{#1}%
\providecommand \bibfnamefont [1]{#1}%
\providecommand \citenamefont [1]{#1}%
\providecommand \href@noop [0]{\@secondoftwo}%
\providecommand \href [0]{\begingroup \@sanitize@url \@href}%
\providecommand \@href[1]{\@@startlink{#1}\@@href}%
\providecommand \@@href[1]{\endgroup#1\@@endlink}%
\providecommand \@sanitize@url [0]{\catcode `\\12\catcode `\$12\catcode
  `\&12\catcode `\#12\catcode `\^12\catcode `\_12\catcode `\%12\relax}%
\providecommand \@@startlink[1]{}%
\providecommand \@@endlink[0]{}%
\providecommand \url  [0]{\begingroup\@sanitize@url \@url }%
\providecommand \@url [1]{\endgroup\@href {#1}{\urlprefix }}%
\providecommand \urlprefix  [0]{URL }%
\providecommand \Eprint [0]{\href }%
\providecommand \doibase [0]{https://doi.org/}%
\providecommand \selectlanguage [0]{\@gobble}%
\providecommand \bibinfo  [0]{\@secondoftwo}%
\providecommand \bibfield  [0]{\@secondoftwo}%
\providecommand \translation [1]{[#1]}%
\providecommand \BibitemOpen [0]{}%
\providecommand \bibitemStop [0]{}%
\providecommand \bibitemNoStop [0]{.\EOS\space}%
\providecommand \EOS [0]{\spacefactor3000\relax}%
\providecommand \BibitemShut  [1]{\csname bibitem#1\endcsname}%
\let\auto@bib@innerbib\@empty
\bibitem [{\citenamefont {Takeuchi}\ \emph {et~al.}(2007)\citenamefont
  {Takeuchi}, \citenamefont {Yasuda}, \citenamefont {Tsujino}, \citenamefont
  {Shishido}, \citenamefont {Settai}, \citenamefont {Harima},\ and\
  \citenamefont {\=Onuki}}]{Takeuchi_2007}%
  \BibitemOpen
  \bibfield  {author} {\bibinfo {author} {\bibfnamefont {T.}~\bibnamefont
  {Takeuchi}}, \bibinfo {author} {\bibfnamefont {T.}~\bibnamefont {Yasuda}},
  \bibinfo {author} {\bibfnamefont {M.}~\bibnamefont {Tsujino}}, \bibinfo
  {author} {\bibfnamefont {H.}~\bibnamefont {Shishido}}, \bibinfo {author}
  {\bibfnamefont {R.}~\bibnamefont {Settai}}, \bibinfo {author} {\bibfnamefont
  {H.}~\bibnamefont {Harima}},\ and\ \bibinfo {author} {\bibfnamefont
  {Y.}~\bibnamefont {\=Onuki}},\ }\href
  {https://doi.org/10.1143/JPSJ.76.014702} {\bibfield  {journal} {\bibinfo
  {journal} {Journal of the Physical Society of Japan}\ }\textbf {\bibinfo
  {volume} {76}},\ \bibinfo {pages} {014702} (\bibinfo {year}
  {2007})}\BibitemShut {NoStop}%
\bibitem [{\citenamefont {Berry}\ \emph {et~al.}(2010)\citenamefont {Berry},
  \citenamefont {Bittar}, \citenamefont {Capan}, \citenamefont {Pagliuso},\
  and\ \citenamefont {Fisk}}]{Berry_2010}%
  \BibitemOpen
  \bibfield  {author} {\bibinfo {author} {\bibfnamefont {N.}~\bibnamefont
  {Berry}}, \bibinfo {author} {\bibfnamefont {E.~M.}\ \bibnamefont {Bittar}},
  \bibinfo {author} {\bibfnamefont {C.}~\bibnamefont {Capan}}, \bibinfo
  {author} {\bibfnamefont {P.~G.}\ \bibnamefont {Pagliuso}},\ and\ \bibinfo
  {author} {\bibfnamefont {Z.}~\bibnamefont {Fisk}},\ }\href
  {https://doi.org/10.1103/PhysRevB.81.174413} {\bibfield  {journal} {\bibinfo
  {journal} {Phys. Rev. B}\ }\textbf {\bibinfo {volume} {81}},\ \bibinfo
  {pages} {174413} (\bibinfo {year} {2010})}\BibitemShut {NoStop}%
\bibitem [{\citenamefont {Cornelius}\ \emph {et~al.}(2001)\citenamefont
  {Cornelius}, \citenamefont {Pagliuso}, \citenamefont {Hundley},\ and\
  \citenamefont {Sarrao}}]{Cornelius_2001}%
  \BibitemOpen
  \bibfield  {author} {\bibinfo {author} {\bibfnamefont {A.~L.}\ \bibnamefont
  {Cornelius}}, \bibinfo {author} {\bibfnamefont {P.~G.}\ \bibnamefont
  {Pagliuso}}, \bibinfo {author} {\bibfnamefont {M.~F.}\ \bibnamefont
  {Hundley}},\ and\ \bibinfo {author} {\bibfnamefont {J.~L.}\ \bibnamefont
  {Sarrao}},\ }\href {https://doi.org/10.1103/PhysRevB.64.144411} {\bibfield
  {journal} {\bibinfo  {journal} {Phys. Rev. B}\ }\textbf {\bibinfo {volume}
  {64}},\ \bibinfo {pages} {144411} (\bibinfo {year} {2001})}\BibitemShut
  {NoStop}%
\bibitem [{\citenamefont {Sheikin}\ \emph {et~al.}(2002)\citenamefont
  {Sheikin}, \citenamefont {Wang}, \citenamefont {Bouquet}, \citenamefont
  {Lejay},\ and\ \citenamefont {Junod}}]{Sheikin_2002}%
  \BibitemOpen
  \bibfield  {author} {\bibinfo {author} {\bibfnamefont {I.}~\bibnamefont
  {Sheikin}}, \bibinfo {author} {\bibfnamefont {Y.}~\bibnamefont {Wang}},
  \bibinfo {author} {\bibfnamefont {F.}~\bibnamefont {Bouquet}}, \bibinfo
  {author} {\bibfnamefont {P.}~\bibnamefont {Lejay}},\ and\ \bibinfo {author}
  {\bibfnamefont {A.}~\bibnamefont {Junod}},\ }\href
  {https://doi.org/10.1088/0953-8984/14/28/104} {\bibfield  {journal} {\bibinfo
   {journal} {Journal of Physics: Condensed Matter}\ }\textbf {\bibinfo
  {volume} {14}},\ \bibinfo {pages} {L543} (\bibinfo {year}
  {2002})}\BibitemShut {NoStop}%
\bibitem [{\citenamefont {Hodovanets}\ \emph {et~al.}(2015)\citenamefont
  {Hodovanets}, \citenamefont {Bud'ko}, \citenamefont {Straszheim},
  \citenamefont {Taufour}, \citenamefont {Mun}, \citenamefont {Kim},
  \citenamefont {Flint},\ and\ \citenamefont {Canfield}}]{Hodovanets_2015}%
  \BibitemOpen
  \bibfield  {author} {\bibinfo {author} {\bibfnamefont {H.}~\bibnamefont
  {Hodovanets}}, \bibinfo {author} {\bibfnamefont {S.~L.}\ \bibnamefont
  {Bud'ko}}, \bibinfo {author} {\bibfnamefont {W.~E.}\ \bibnamefont
  {Straszheim}}, \bibinfo {author} {\bibfnamefont {V.}~\bibnamefont {Taufour}},
  \bibinfo {author} {\bibfnamefont {E.~D.}\ \bibnamefont {Mun}}, \bibinfo
  {author} {\bibfnamefont {H.}~\bibnamefont {Kim}}, \bibinfo {author}
  {\bibfnamefont {R.}~\bibnamefont {Flint}},\ and\ \bibinfo {author}
  {\bibfnamefont {P.~C.}\ \bibnamefont {Canfield}},\ }\href
  {https://doi.org/10.1103/PhysRevLett.114.236601} {\bibfield  {journal}
  {\bibinfo  {journal} {Phys. Rev. Lett.}\ }\textbf {\bibinfo {volume} {114}},\
  \bibinfo {pages} {236601} (\bibinfo {year} {2015})}\BibitemShut {NoStop}%
\bibitem [{\citenamefont {Lai}\ \emph {et~al.}(2018)\citenamefont {Lai},
  \citenamefont {Grefe}, \citenamefont {Paschen},\ and\ \citenamefont
  {Si}}]{lai2018weyl}%
  \BibitemOpen
  \bibfield  {author} {\bibinfo {author} {\bibfnamefont {H.-H.}\ \bibnamefont
  {Lai}}, \bibinfo {author} {\bibfnamefont {S.~E.}\ \bibnamefont {Grefe}},
  \bibinfo {author} {\bibfnamefont {S.}~\bibnamefont {Paschen}},\ and\ \bibinfo
  {author} {\bibfnamefont {Q.}~\bibnamefont {Si}},\ }\href
  {https://doi.org/10.1073/pnas.1715851115} {\bibfield  {journal} {\bibinfo
  {journal} {Proceedings of the National Academy of Sciences}\ }\textbf
  {\bibinfo {volume} {115}},\ \bibinfo {pages} {93} (\bibinfo {year}
  {2018})}\BibitemShut {NoStop}%
\bibitem [{\citenamefont {Kushwaha}\ \emph {et~al.}(2019)\citenamefont
  {Kushwaha}, \citenamefont {Chan}, \citenamefont {Park}, \citenamefont
  {Thomas}, \citenamefont {Bauer}, \citenamefont {Thompson}, \citenamefont
  {Ronning}, \citenamefont {Rosa},\ and\ \citenamefont
  {Harrison}}]{Kushwaha2019}%
  \BibitemOpen
  \bibfield  {author} {\bibinfo {author} {\bibfnamefont {S.~K.}\ \bibnamefont
  {Kushwaha}}, \bibinfo {author} {\bibfnamefont {M.~K.}\ \bibnamefont {Chan}},
  \bibinfo {author} {\bibfnamefont {J.}~\bibnamefont {Park}}, \bibinfo {author}
  {\bibfnamefont {S.~M.}\ \bibnamefont {Thomas}}, \bibinfo {author}
  {\bibfnamefont {E.~D.}\ \bibnamefont {Bauer}}, \bibinfo {author}
  {\bibfnamefont {J.~D.}\ \bibnamefont {Thompson}}, \bibinfo {author}
  {\bibfnamefont {F.}~\bibnamefont {Ronning}}, \bibinfo {author} {\bibfnamefont
  {P.~F.~S.}\ \bibnamefont {Rosa}},\ and\ \bibinfo {author} {\bibfnamefont
  {N.}~\bibnamefont {Harrison}},\ }\href
  {https://doi.org/10.1038/s41467-019-13421-w} {\bibfield  {journal} {\bibinfo
  {journal} {Nature Communications}\ }\textbf {\bibinfo {volume} {10}},\
  \bibinfo {pages} {5487} (\bibinfo {year} {2019})}\BibitemShut {NoStop}%
\bibitem [{\citenamefont {van Dijk}\ \emph {et~al.}(2000)\citenamefont {van
  Dijk}, \citenamefont {F\aa{}k}, \citenamefont {Charvolin}, \citenamefont
  {Lejay},\ and\ \citenamefont {Mignot}}]{Dijk_2000}%
  \BibitemOpen
  \bibfield  {author} {\bibinfo {author} {\bibfnamefont {N.~H.}\ \bibnamefont
  {van Dijk}}, \bibinfo {author} {\bibfnamefont {B.}~\bibnamefont {F\aa{}k}},
  \bibinfo {author} {\bibfnamefont {T.}~\bibnamefont {Charvolin}}, \bibinfo
  {author} {\bibfnamefont {P.}~\bibnamefont {Lejay}},\ and\ \bibinfo {author}
  {\bibfnamefont {J.~M.}\ \bibnamefont {Mignot}},\ }\href
  {https://doi.org/10.1103/PhysRevB.61.8922} {\bibfield  {journal} {\bibinfo
  {journal} {Phys. Rev. B}\ }\textbf {\bibinfo {volume} {61}},\ \bibinfo
  {pages} {8922} (\bibinfo {year} {2000})}\BibitemShut {NoStop}%
\bibitem [{\citenamefont {Knafo}\ \emph {et~al.}(2003)\citenamefont {Knafo},
  \citenamefont {Raymond}, \citenamefont {Fåk}, \citenamefont {Lapertot},
  \citenamefont {Canfield},\ and\ \citenamefont {Flouquet}}]{WKnafo_2003}%
  \BibitemOpen
  \bibfield  {author} {\bibinfo {author} {\bibfnamefont {W.}~\bibnamefont
  {Knafo}}, \bibinfo {author} {\bibfnamefont {S.}~\bibnamefont {Raymond}},
  \bibinfo {author} {\bibfnamefont {B.}~\bibnamefont {Fåk}}, \bibinfo {author}
  {\bibfnamefont {G.}~\bibnamefont {Lapertot}}, \bibinfo {author}
  {\bibfnamefont {P.~C.}\ \bibnamefont {Canfield}},\ and\ \bibinfo {author}
  {\bibfnamefont {J.}~\bibnamefont {Flouquet}},\ }\href
  {https://doi.org/10.1088/0953-8984/15/22/308} {\bibfield  {journal} {\bibinfo
   {journal} {Journal of Physics: Condensed Matter}\ }\textbf {\bibinfo
  {volume} {15}},\ \bibinfo {pages} {3741} (\bibinfo {year}
  {2003})}\BibitemShut {NoStop}%
\bibitem [{\citenamefont {Simeth}\ \emph {et~al.}(2023)\citenamefont {Simeth},
  \citenamefont {Wang}, \citenamefont {Ghioldi}, \citenamefont {Fobes},
  \citenamefont {Podlesnyak}, \citenamefont {Sung}, \citenamefont {Bauer},
  \citenamefont {Lass}, \citenamefont {Flury}, \citenamefont {Vonka},
  \citenamefont {Mazzone}, \citenamefont {Niedermayer}, \citenamefont {Nomura},
  \citenamefont {Arita}, \citenamefont {Batista}, \citenamefont {Ronning},\
  and\ \citenamefont {Janoschek}}]{simeth2022microscopic}%
  \BibitemOpen
  \bibfield  {author} {\bibinfo {author} {\bibfnamefont {W.}~\bibnamefont
  {Simeth}}, \bibinfo {author} {\bibfnamefont {Z.}~\bibnamefont {Wang}},
  \bibinfo {author} {\bibfnamefont {E.~A.}\ \bibnamefont {Ghioldi}}, \bibinfo
  {author} {\bibfnamefont {D.~M.}\ \bibnamefont {Fobes}}, \bibinfo {author}
  {\bibfnamefont {A.}~\bibnamefont {Podlesnyak}}, \bibinfo {author}
  {\bibfnamefont {N.~H.}\ \bibnamefont {Sung}}, \bibinfo {author}
  {\bibfnamefont {E.~D.}\ \bibnamefont {Bauer}}, \bibinfo {author}
  {\bibfnamefont {J.}~\bibnamefont {Lass}}, \bibinfo {author} {\bibfnamefont
  {S.}~\bibnamefont {Flury}}, \bibinfo {author} {\bibfnamefont
  {J.}~\bibnamefont {Vonka}}, \bibinfo {author} {\bibfnamefont {D.~G.}\
  \bibnamefont {Mazzone}}, \bibinfo {author} {\bibfnamefont {C.}~\bibnamefont
  {Niedermayer}}, \bibinfo {author} {\bibfnamefont {Y.}~\bibnamefont {Nomura}},
  \bibinfo {author} {\bibfnamefont {R.}~\bibnamefont {Arita}}, \bibinfo
  {author} {\bibfnamefont {C.~D.}\ \bibnamefont {Batista}}, \bibinfo {author}
  {\bibfnamefont {F.}~\bibnamefont {Ronning}},\ and\ \bibinfo {author}
  {\bibfnamefont {M.}~\bibnamefont {Janoschek}},\ }\href
  {https://doi.org/10.1038/s41467-023-43947-z} {\bibfield  {journal} {\bibinfo
  {journal} {Nature Communications}\ }\textbf {\bibinfo {volume} {14}},\
  \bibinfo {pages} {8239} (\bibinfo {year} {2023})}\BibitemShut {NoStop}%
\end{thebibliography}

%


\newpage



{
\section*{Supplemental Information for Excess heat capacity in magnetically ordered Ce heavy fermion metals}

\renewcommand{\thefigure}{S\arabic{figure}}
\renewcommand{\thetable}{S\arabic{table}}
\renewcommand{\theequation}{S.\arabic{equation}}
\renewcommand{\thepage}{S\arabic{page}}  
\setcounter{figure}{0}
\setcounter{page}{1}
\setcounter{equation}{0}

\section{Heat capacity fits}

The magnetic heat capacities and $c = \gamma T + N_a k_B \big(\frac{T}{T_{\beta}} \big)^3$ 
fits used to populate Table I of the main text are shown in Fig. \ref{fig:HCmagnons}. 
For each compound, we fitted the lowest temperature section of data to estimate the $T^3$ behavior. There are two exceptions to this: CeCu$_2$Ge$_2$ (panel h) and CePt$_3$Si (panel k). For  CeCu$_2$Ge$_2$, there is some ambiguity as to the fitted region: 
fitting the data below 0.1~K (shown with a dashed line) gives $T_{\beta} = 2.08(6)$~K, whereas fitting the data above 0.2~K (where a $T^3$ region is more apparent) gives $T_{\beta} = 3.51(2)$~K 
(as shown in Fig. \ref{fig:ModCorrPlot}, neither choice gives any overall correlation between $P_c$, which is the key result of this exercise).  For CePt$_3$Si, there is an additional superconducting transition at $T_c = 0.46$~K \cite{Takeuchi_2007}, and we discard the data below this as it no longer represents magnon specific heat. 
For the five compounds plotted in main text Fig. 2, we subtracted the specific heat of a nonmagnetic La analogue:  LaIn$_3$ \cite{Berry_2010}, LaRhIn$_5$ \cite{Cornelius_2001}, LaPd$_2$Si$_2$ \cite{Sheikin_2002}, LaCu$_2$Ge$_2$ \cite{Hodovanets_2015}, and LaPt$_3$Si \cite{Takeuchi_2007}.

As is immediately obvious, not all of these compounds display clear $T^3$ behavior. However, all of them do have some significant slope at the lowest temperature, indicating more than just a constant Sommerfeld offset. 
Although the choice of where to fit $T_{\beta}$ is somewhat arbitrary, it is clear that excess temperature dependent heat capacity is a common feature in the heavy fermion magnets.

\begin{figure*}
	\centering
	\includegraphics[width=0.9\textwidth]{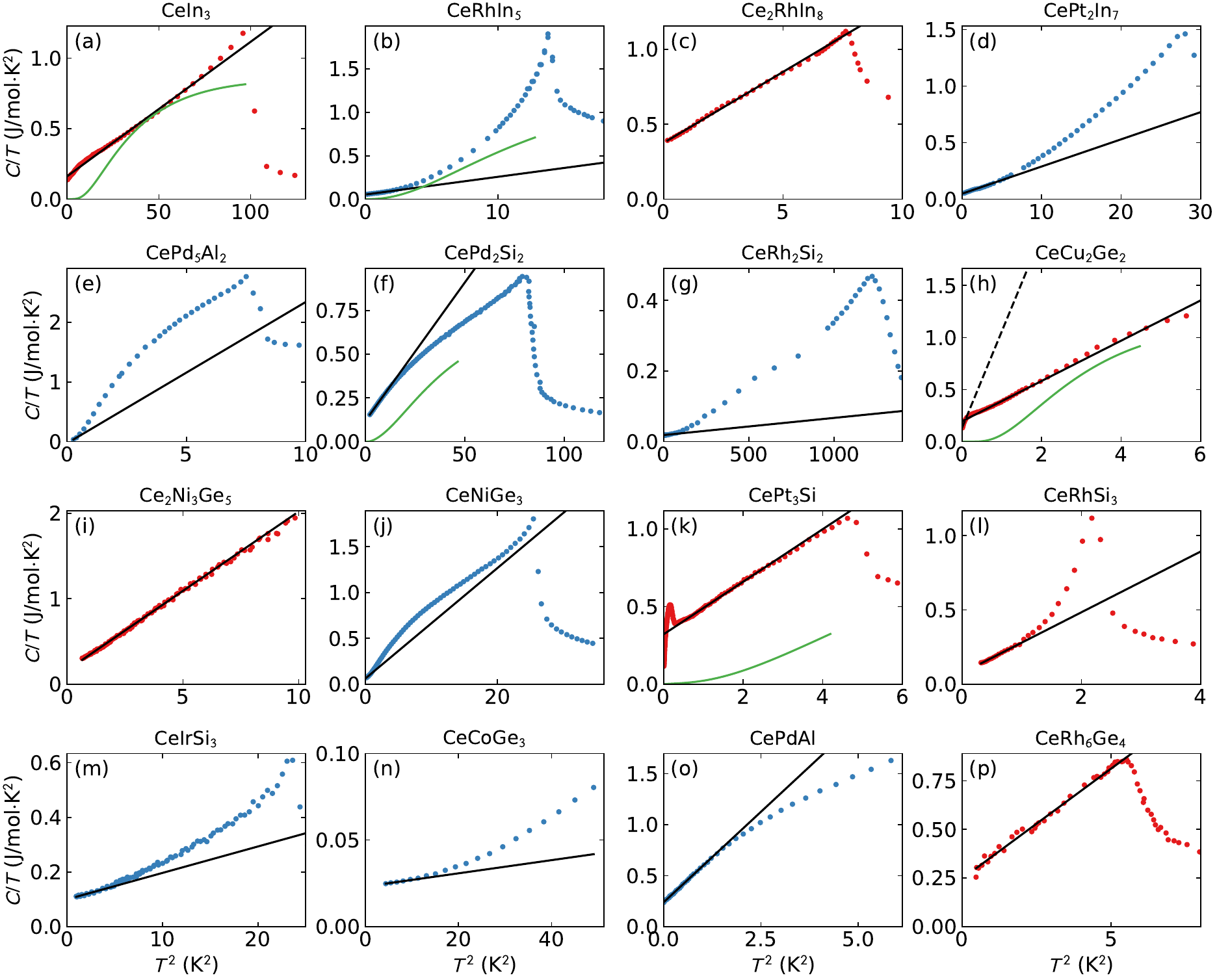}
	\caption{Heat capacity of 16 magnetically ordered heavy fermion materials. The data plotted in red have a $T^3$ behavior below $T_N$ over a sizeable region. The black line is the $T^3$ fit, and the green lines show the heat capacity from the fitted magnon model (where available).}
	\label{fig:HCmagnons}
\end{figure*}

\begin{figure}
	\centering
	\includegraphics[width=0.5\textwidth]{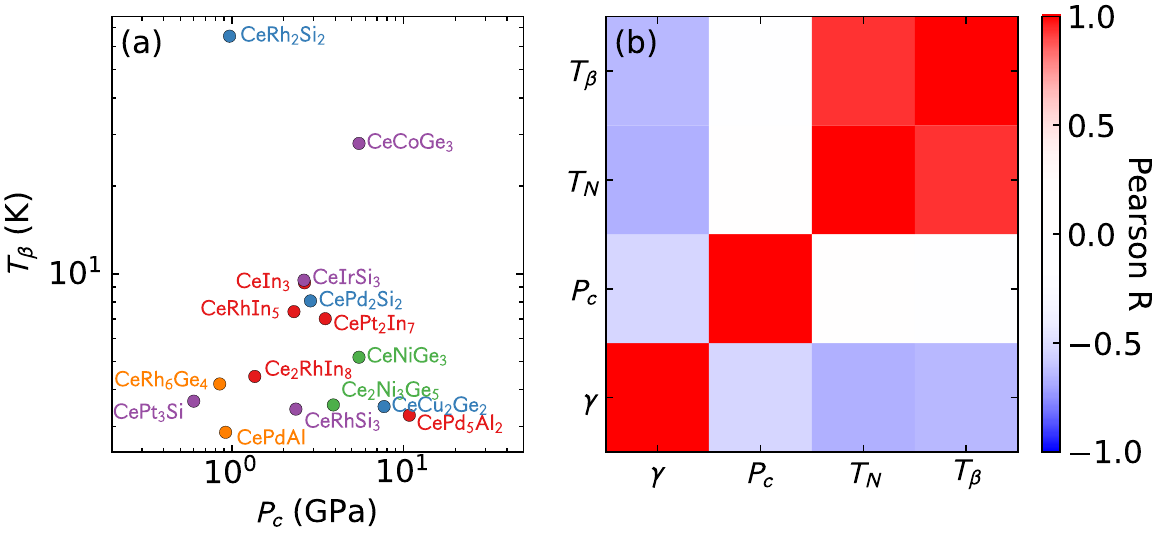}
	\caption{(a) Correlation plot between $T_{\beta}$ and $P_c$ assuming the larger CeCu$_2$Ge$_2$ $\beta = 0.93(8)$~mJ/mol$\cdot$K$^4$. Panel (b) shows the recalculated correlation matrix, which is little changed from the main text Fig. 3.}
	\label{fig:ModCorrPlot}
\end{figure}

Figure \ref{fig:LinearCorrelation} shows the compounds in main text Table I with $T_{\beta}$ plotted against $T_N$, along with a power law and linear fit. These two temperature scales have a nearly linear relationship, and are even of the same order of magnitude. This is a strong indication that the magnetic exchange energy is the key to explaining the anomalous low temperature density of states in Ce heavy fermions. 

\begin{figure}
	\centering
	\includegraphics[width=0.44\textwidth]{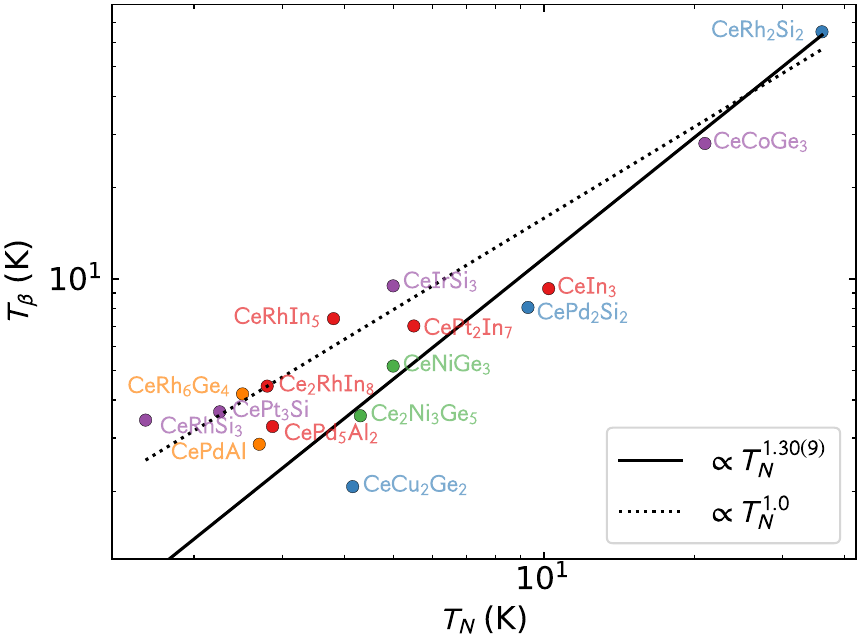}
	\caption{Correlation plot between $T_{\beta}$ and $T_N$, along with a power law fit (solid line) and a linear fit (dashed line). The fitted power law is nearly 1, and the linear fit matches the data reasonably well.}
	\label{fig:LinearCorrelation}
\end{figure}

For the five compounds in main text Fig. 2, the residual heat capacity resembles a gapped density of states at the lowest energies. In Table \ref{tab:residualHC}, we enumerate various characteristics of this (pseudo)gap: the entropy, rise in heat capacity, and estimated gap. These estimates provide a rough picture of the excess density of states across the five materials. 
Interestingly, the pseudogap $\Delta$ is of the same order as $T_N$ for these materials, ranging from $0.7 T_N k_B$ for CeCu$_2$Ge$_2$ to $2.9 T_N k_B$ for CePt$_3$Si. This is consistent with the energy scale being governed by magnetic exchange interactions. 

\begin{table}
\caption{Characteristics of the magnon-model-subtracted specific heat of the five compounds in main text Fig. 2. We show entropy to $T_N/2$ ($\Delta S$), the magnitude rise in $C/T$ above $\gamma$ (Pseudogap $C_e/T$), and the gap from a Schottky comparison (Pseudogap $\Delta$).}
\begin{ruledtabular}
\begin{tabular}{c|ccc}
Compound & $\Delta S$ to $T_N/2$ & Pseudogap $C_e/T$ & Pseudogap $\Delta$ \\
  &  (J/mol$\cdot$K) & (J/mol$\cdot$K$^2$) &  (meV) \\
\hline
CeIn$_3$ & 0.857 & 0.098 & 0.88 \\
CeRhIn$_5$ & 0.113 & 0.012 & 0.41 \\
CePd$_2$Si$_2$ & 0.503 & 0.080 & 1.15 \\
CeCu$_2$Ge$_2$ & 0.605 & 0.174 & 0.25 \\
CePt$_3$Si & 0.494 & 0.264 & 0.57 \\
\end{tabular}\end{ruledtabular}
\label{tab:residualHC}
\end{table}

\section{Correlation Analysis}

Figure \ref{fig:CorrelationPoints} shows the correlation plots used to create the correlation matrix in main text Fig. 3. As can be seen, some variables (like $\beta$ and $T_N$) are strongly correlated, while others are not.

\begin{figure}
	\centering
	\includegraphics[width=0.48\textwidth]{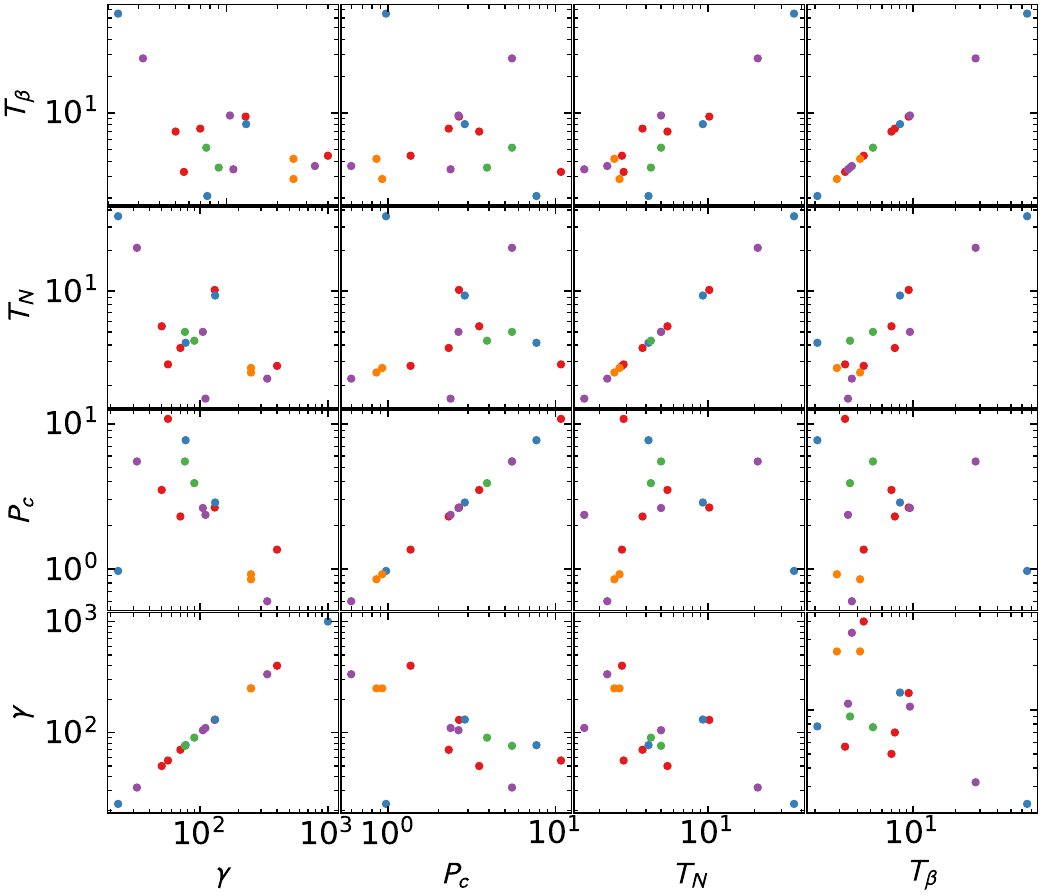}
	\caption{Correlation plots used to generate main text Fig. 3, plotted on a log-log scale. }
	\label{fig:CorrelationPoints}
\end{figure}

We can take this correlation analysis a step further by explicitly analyzing the variance in $T_{\beta}$ which can be accounted for by the other variables. 
We do this by fitting the equation 
\begin{equation}
    T^{\rm calc}_{\beta} = a_1 \gamma + a_2 P_c + a_3 T_N \label{eq:LinearFit}
\end{equation}
to the experimental $T_{\beta}$, and then systematically eliminating variables from Eq. \ref{eq:LinearFit} and refitting the $a_i$ parameters. We then computing the residual variance $var(T^{\rm calc}_{\beta} - T_{\beta})$, which reveals what fraction of the variance can be explained by various combinations of variables. To account for nonlinear correlations, we also fit a nonlinear equation allowing for power law correlations
\begin{equation}
    T^{\rm calc}_{\beta} = a_1 \gamma^{b_1} + a_2 P_c^{b_2}  + a_3 T_N^{b_3}  \label{eq:NonLinearFit}
\end{equation}
where $a_i$ and $b_i$ are fitted freely.
 The results of both sets of fits are shown in Table \ref{tab:variances}. 

\begin{table}
\caption{Residual variances of fits to Eq. \ref{eq:LinearFit} (linear) and Eq. \ref{eq:NonLinearFit} (nonlinear). The variance explained is the ratio between the residual variance of the fit and the raw variance of $T_{\beta}$. For both the linear and nonlinear fits, the variance is almost entirely accounted for by $T_N$, and adding other variables to the fit decreases the variance by less than a percent. }
\begin{ruledtabular}
\begin{tabular}{c|cc}
variables & \% var. explained (linear) & \% var. explained (nonlinear) \\
\hline
$\gamma$        & 15.84 & 40.95 \\
$P_c$           & 2.60 & 2.60 \\
$T_N$           & 96.51 & 98.65 \\
$\gamma$,$P_c$  & 35.12 & 78.81 \\
$\gamma$,$T_N$  & 96.88 & 98.67 \\
$P_c$,$T_N$     & 96.61 & 98.77 \\
$\gamma$,$P_c$,$T_N$ & 96.88 & 99.18
\end{tabular}\end{ruledtabular}
\label{tab:variances}
\end{table}

The fits in Table \ref{tab:variances} reveal that more than 96\% of the $T_{\beta}$ variance can be accounted for by $T_N$ (more than 98\% with a nonlinear fit). Although there is nonzero variance explained by $\gamma$ in isolation, combining this with $T_N$ reveals a negligible difference in residual variance as compared to $T_N$ alone, which means that the variance explained \textit{independent of $T_N$} is $< 0.4$~\% for $\gamma$ and $< 0.1$~\% for $P_c$. 
Therefore, despite nonzero Pearson R correlations in the matrix in the main text Fig. 4, the variance in $T_{\beta}$ is almost entirely due to $T_N$. 

\section{Dirac cone specific heat}

As mentioned in the main text, a common explanation for a $T^3$ specific heat is a linear Dirac cone dispersion where the crossing is near the Fermi energy. (In this context, "Dirac Cone" simply means a linear dispersing mode as in the Dirac equation, and does not imply nontrivial topology.) Generically, the integral 
$$
    c_v = k_B \sum_s \int dk  \Big( \frac{\hbar \omega_s (k)}{k_B T} \Big)^2 \frac{e^{\hbar \omega_s (k) / k_B T}}{(e^{\hbar \omega_s (k) / k_B T} +1)^2}
$$
can equivalently be written as an energy integral over the density of states $D(\epsilon)$
\begin{equation}
    c_v = \frac{1}{k_B T^2} \int d \epsilon \> D(\epsilon) \Big( \frac{\epsilon}{e^{\epsilon / k_B T} +1} \Big)^2 e^{\epsilon / k_B T}.
\end{equation}
Assuming two fermionic $\omega = ck$ bands yields the density of states $D(\epsilon) = \int dk \> \delta \big(\epsilon - \omega(k) \big) = \frac{\epsilon^2}{\pi^2 (\hbar c)^3}$ and specific heat
\begin{equation}
c_v = \frac{7 k_B^4 \pi^2 T^3}{30(\hbar c)^3}
\label{eq:ZeroDiracHC}
\end{equation}
\cite{lai2018weyl}. However, if we allow the Fermi surface to deviate from the Dirac crossing such that $\omega = ck - \epsilon_f$, the corresponding three-dimensional density of states is $D(\epsilon) = \frac{1}{\pi^2} \frac{(\epsilon - \epsilon_f)^2}{(\hbar c)^3}$ and the derived equation is
\begin{equation}
    c_v = \frac{k_B^2}{30 (\hbar c)^3 \pi^2} \big(5 \epsilon_f^2 \pi^2 T - 270 \epsilon_f k_B \zeta(3) T^2 + 7 k_B^2 \pi^4 T^3 \big)
\end{equation}
where $\zeta(3)$ is the Riemann zeta function. In this case, a nonzero $\epsilon_f$ yields a finite $c_v/T$ as $T \rightarrow 0$ and a $T^2$ term, but the $T^3$ term is unchanged as $\epsilon_f$ deviates from the crossing energy. Thus approximate $T^3$ behavior will be preserved if $\epsilon_f$ is small. However, this hypothesis has two difficulties when it comes to CeIn$_3$ and related compounds:  (i) there is no clear reason why the velocity should correlate with $T_N$, and (ii) it assumes that every compound studied here happens to have a linear crossing close to the Fermi energy, for which no mechanism has been proposed. 

\subsection*{Pseudogap model}

In the main text we compare the psudogap behavior to a Schottky anomaly, but this is by no means the only density of states one could postulate. For instance, a gapped Schotte-Schotte curve \cite{Kushwaha2019} would also resemble the data, as would a variety of gapped distributions. However, we stress that  phenomenological fits to a single heat capacity curve is insufficient for precisely determining the underlying model. Our aim is to point out the need for a microscopic model. 

\section{$\rm CePd_2Si_2$ magnon model}

Although the magnon model for CePd$_2$Si$_2$ has been fitted to neutron scattering data in Ref. \cite{Dijk_2000}, there is room for skepticism because the experiments were performed by triple axis neutron scattering. As the example of CeIn$_3$ shows, triple axis measurements may indicate a gapped magnon spectrum \cite{WKnafo_2003} whereas higher-resolution measurements show a gapless spectrum  \cite{simeth2022microscopic}. 
We would therefore not be surprised if CePd$_2$Si$_2$ is like CeIn$_3$ in that it is actually gapless but its steep dispersion was simply not resolvable in the experiments performed. 

To account for this, we take the model presented in Ref. \cite{Dijk_2000}, set the spin gap to zero, and recalculate the heat capacity as shown in Fig. \ref{fig:CePd2Si2}. Because the bottom of the mode takes up only a small fraction of the Brillouin zone and the magnon bandwidth is small, this make very little difference in the calculated heat capacity. Indeed, the only difference is a slight increase of slope at the lowest temperatures, but this slope is still smaller than experiment by a factor of 2.8. 
Therefore the discrepancy between experimental heat capacity and the magnon modeled heat capacity for CePd$_2$Si$_2$ cannot be resolved by assuming gapless modes. 

\begin{figure*}
	\centering
	\includegraphics[width=0.8\textwidth]{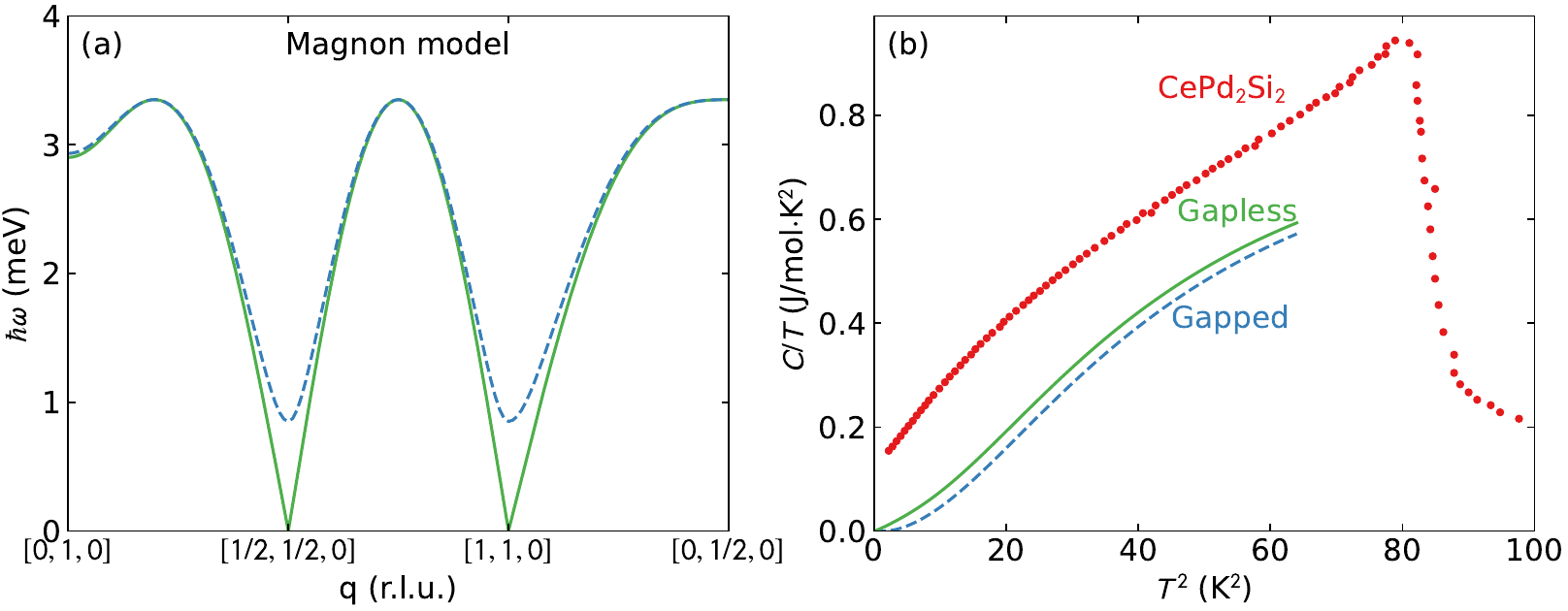}
	\caption{Heat capacity computed from magnon models for CePd$_2$Si$_2$. (a) Gapped magnon model from Ref. \cite{Dijk_2000} (dashed blue) and the same model modified to have zero anisotropy and thus gapless (green). (b) CePd$_2$Si$_2$ heat capacity from Ref. \cite{Sheikin_2002} compared with the magnon specific heat of the two models. Besides marginally more density of states at low temperature, the two curves are identical, and do not match the slope of the experimental specific heat as $T \rightarrow 0$. }
	\label{fig:CePd2Si2}
\end{figure*}

}

\end{document}